\documentclass[twocolumn,showpacs,superscriptaddress,amsmath,amssymb]{revtex4}
\usepackage{graphicx}
\usepackage{dcolumn}
\usepackage{bm}
\usepackage{color}
\usepackage{mathrsfs}
\usepackage{ulem}
\usepackage[german]{babel}

\def\bd{
\begin{document}} \def\ed{\end{document}}
\def\bmp{\begin{minipage}} \def\emp{\end{minipage}}
\def\bcc{\begin{center}} \def\ecc{\end{center}}     \def\npg{\newpage}
\def\beq{\begin{equation}} \def\eeq{\end{equation}} \def\hph{\hphantom}
\def\be{\begin{equation}} \def\ee{\end{equation}} \def\r#1{$^{[#1]}$}
\def\n{\noindent} \def\ni{\noindent} \def\pa{\parindent}
\def\hs{\hskip} \def\vs{\vskip} \def\hf{\hfill} \def\ej{\vfill\eject}
\def\cl{\centerline} \def\ob{\obeylines}  \def\ls{\leftskip}
\def\underbar#1{$\setbox0=\hbox{#1} \dp0=1.5pt \mathsurround=0pt
   \underline{\box0}$}   \def\ub{\underbar}    \def\ul{\underline}
\def\f{\left} \def\g{\right} \def\e{{\rm e}} \def\o{\over} \def\d{{\rm d}}
\def\vf{\varphi} \def\pl{\partial} \def\cov{{\rm cov}} \def\ch{{\rm ch}}
\def\la{\langle} \def\ra{\rangle} \def\EE{e$^+$e$^-$} \def\pt{p_{\rm t}}
\def\pti{p_{{\rm t},i}} \def\vti{v_{{\rm t},i}}
\def\ptj{p_{{\rm t},j}}\def\Pt{P_{\rm t}} \def\vt{v_{\rm t}}

\def\bitz{\begin{itemize}} \def\eitz{\end{itemize}}
\def\btbl{\begin{tabular}} \def\etbl{\end{tabular}}
\def\btbb{\begin{tabbing}} \def\etbb{\end{tabbing}}
\def\beqar{\begin{eqnarray}} \def\eeqar{\end{eqnarray}}
\def\\{\hfill\break} \def\dit{\item{-}} \def\i{\item}
\def\bbb{\begin{thebibliography}{9} \itemsep=-1mm}
\def\ebb{\end{thebibliography}} \def\bb{\bibitem}
\def\bpic{\begin{picture}(260,240)} \def\epic{\end{picture}}
\def\akgt{\cl{\bf ACKNOWLEDGMENTS}}
\def\fgn{\noindent{\bf\large\bf figure captions}}
\def\m1{\langle N_p\rangle} \def\u2{\langle N_{\bar p}\rangle} \def\Nap{N_{\bar
p}}
\def\lan{\langle}
\def\ran{\rangle}
\def\p{\pi}
\def\ifmath#1{\relax\ifmmode #1\else $#1$\fi}%
\def\rc{\ifmath{{\mathrm{c}}}}
\def\cut{\ifmath{{\mathrm{cut}}}}
\def\rF{\ifmath{{\mathrm{F}}}}
\def\rK{\ifmath{{\mathrm{K}}}}
\def\rp{\ifmath{{\mathrm{p}}}}
\def\rt{\ifmath{{\mathrm{t}}}}
\def\LAB{\ifmath{{\mathrm{LAB}}}}
\def\cut{\ifmath{{\mathrm{cut}}}}
\def\beq{\begin{equation}}
\def\eeq{\end{equation}}

\newcommand{\cinst}[2]{$^{\mathrm{#1}}$~#2\par}
\newcommand{\crefi}[1]{$^{\mathrm{#1}}$}
\newcommand{\crefii}[2]{$^{\mathrm{#1,#2}}$}
\newcommand{\crefiii}[3]{$^{\mathrm{#1,#2,#3}}$}
\newcommand{\HRule}{\rule{0.5\linewidth}{0.5mm}}

\bd
\title{Fixed point behavior mapping of cumulants between the three-dimensional Ising model and QCD}

\author{Xue Pan}\email{panxue1624@163.com}
\affiliation{School of Electronic Information and Electrical Engineering, \\ Chengdu University, Chengdu 610106, China}

\begin{abstract}
Fixed point behavior was found in the temperature dependence of normalized cumulants of order parameter at different external magnetic fields in the three-dimensional Ising model in my last work. In this paper, considering possible existing QCD critical point belonging to the three-dimensional Ising universality class and the non-universal mapping parameters between the Ising model and QCD, effects of the mapping parameters on the fixed point behavior in the net-baryon chemical potential dependence of normalized cumulants along different experimental freeze-out curves is studied in this paper. We found that when the directions of Ising variables, the reduced temperature and external magnetic field, are orthogonal or not far from orthogonal after mapping to the QCD temperature and net-baryon chemical potential plane, the fixed point behavior exists in the net-baryon chemical potential dependence of the normalized cumulants and is closer to the QCD critical point than the peak in the cumulants.
\end{abstract}


\maketitle

\section{Introduction}

One of the main goals of current relativistic heavy-ion collision experiments is to reveal the phase diagram of quantum chromo-dynamics (QCD)~\cite{maingoal}, where the location of the critical point, which is a unique character of the QCD phase diagram, is the most important.
Recent years, high-order cumulants of multiplicity distributions of conserved charges, such as net-baryon, net-charge and net-strangeness, are suggested to search for the QCD critical point~\cite{stephanov-prl91, koch, Stephanov-prl102, Karsch-EPJC71}.

In the vicinity of the critical point, non-monotonic behavior of the high-order cumulants of conserved charges are found in a variety of QCD effective models and is regarded as critical related signal~\cite{Stephanov-prl102, Asakawa-prl103, Fuweijie, Vladi}. Particularly, when the critical point is approached on the crossover side, universal negative fourth-order cumulant of the order parameter is predicted in Ref.~\cite{Stephanov-prl107} and used to locate the QCD critical point in experiments~\cite{Phys.Rev.Lett.112.032302}. While in Refs.~\cite{Chin.Phys.C.43.033103,Chin.Phys.C.45.104103}, the authors argued that the sign of high-order cumulants is not sufficient to prove the presence of the critical point. Other work indicate that the peak structure of kurtosis remains a solid feature and can be used as a clean signature of the critical point~\cite{Eur.Phys.J.C.79.245, Phys.Rev.C.103.034901}. Nevertheless the peak structure of cumulants can also occurs in a first order phase transition or crossover~\cite{Nature,JPG}. Whether the non-monotonic or sign change behavior of high-order cumulants of conserved charges is reliable, accurate or enough to locate the critical point is uncertain.

In fact, the finite-size scaling, which implies a fixed point, can also be employed to locate the critical point~\cite{Phys.Rev.D.97.034015, JPG, Phys.Rev.E.100.052146, chenlz}. Usually, the fixed point is obtained from the scale transformation of the re-normalization group, resulting in the independence of rescaled thermodynamics on the system sizes at the critical point~\cite{fixedpoint2, fixedpoint3}.

In my recent work, the fixed point behavior has been found in the temperature dependence of normalized high-order cumulants of the order parameter at different fixed external magnetic fields in the three-dimensional Ising model~\cite{cpc2022}. What is more, the fixed point is just at the critical temperature of the Ising model.

The QCD critical point, if exists, is expected to belong to the same universality class of the three-dimensional Ising model~\cite{class1, class2, class3, class4}. Critical behavior of the corresponding thermodynamics in different systems that belongs to the same universality class is the same which is supervised by the same critical exponents. Recently, many works have been done to map the results of the three-dimensional Ising model to that of the QCD~\cite{PhysRevD102014505, Phys.Rev.C.103.034901}.

In order to map the results of the Ising model to that of the QCD, usually, a linear ansatz between the QCD variables, temperature $T$ and net-baryon chemical potential $\mu_B$, and the Ising variables, reduced temperature $t$ ($t=(\tau-\tau_c)/\tau_c$, where $\tau$ and $\tau_c$ are the temperature and critical temperature of the three-dimensional Ising model, respectively) and external magnetic field $h$ is suggested~\cite{linearmap1, linearmap2, linearmap3, NPA}. Using three assumptions, firstly, the QCD system is in equilibrium, secondly, the linear map keep Ising $t$-direction and $h$-direction orthogonal to each other after mapped to the QCD $T-\mu_B$ phase diagram. At last, the freeze-out curves (FC) are below the crossover/first-order phase transition line, and parallel to the $t$-direction which has been mapped to the QCD phase diagram, the fixed point behavior has been found in the energy (transformation from the net-baryon chemical potential) dependence of normalized cumulants of order parameter in Ref.~\cite{cpc2022}. The fixed point is just at the critical energy (corresponding to the net-baryon chemical potential at the QCD critical point).

However, the mapping parameters are non-universal and far unknown from the Ising model to QCD. In this paper, on the assumption of the equilibrium of the QCD system, using parametric representation of the three-dimensional Ising model, we study and discuss the influences of the mapping parameters on the fixed point behavior in the net-baryon chemical potential dependence of normalized cumulants of order parameter along the experimental freeze-out curves. The freeze-out curves are also supposed below the crossover/first-order phase transition line, but are described by an empirical parametrization of the heavy-ion-collision data~\cite{Phys.Rev.C.73.034905}.

The paper is organized as follows. In section 2, the parametric representation and corresponding parametric expressions of second- to fourth-order cumulants of the three-dimensional Ising model is presented. The linear mapping from the three-dimensional Ising model to QCD is discussed. The experimental freeze-out curves are mapped to the Ising $t-h$ plane through different parameters. In section 3, net-baryon chemical potential dependence of the second- to fourth-order normalized cumulants is studied along the freeze-out curves. The effects of different mapping parameters on the net-baryon chemical potential dependence of the normalized cumulants are discussed. Finally, conclusions and summary are given in section 4.

\section{The linear mapping from Ising to QCD}

In the parametric representation of the three-dimensional Ising model, magnetization $M$ and reduced temperature $t$ can be parameterized by two variables $R$ and $\theta$~\cite{linearpara, linearpara3},
\begin{equation}\label{parametric}
M=m_0R^{\beta}\theta,~~~~~~t=R(1-\theta^2).
\end{equation}
The equation of state expressed by $R$ and $\theta$ is
\begin{equation}\label{equation state}
h=h_0R^{\beta\delta}\widetilde{h}(\theta).
\end{equation}
Where $m_0$ in Eq.~\eqref{parametric} and $h_0$ in Eq.~\eqref{equation state} are normalization constants. They are fixed by imposing the normalization conditions $M(t=-1,h=+0)=1$ and $M(t=0,h=1)=1$. $\beta$ and $\delta$ are critical exponents of the three-dimensional Ising universality class with values 0.3267(10) and 4.786(14), respectively~\cite{Isingexponents}.

One simple mean-field approximation of representation for $\widetilde{h}(\theta)$ is as follows,
\begin{equation}\label{equation h}
\widetilde{h}(\theta)=\theta(3-2\theta^{2}).
\end{equation}

When taking the approximate values of the critical exponents $\beta=1/3$ and $\delta=5$ (it is enough for our purpose), the first fourth order cumulants of magnetization (order parameter of the three dimensional Ising model) in the parametric representation are as follows,
\begin{equation}\label{first four cumulants}
\begin{split}
&\kappa_{1}(R,\theta)=m_0R^{1/3}\theta,\\
&\kappa_{2}(R,\theta)=\frac{m_0}{h_0}\frac{1}{R^{4/3}(2\theta^2+3)},\\
&\kappa_{3}(R,\theta)=\frac{m_0}{h_0^2}\frac{4\theta(\theta^2+9)}{R^{3}(\theta^2-3)(2\theta^2+3)^3},\\
&\kappa_{4}(R,\theta)=12\frac{m_0}{h_0^3}\frac{(2\theta^8-5\theta^6+105\theta^4-783\theta^2+81)}{R^{14/3}(\theta^2-3)^3(2\theta^2+3)^5}.\\
\end{split}
\end{equation}

The reduced temperature $t$ and external magnetic field $h$, are functions of $R$ and $\theta$ provided by Eq.~\eqref{parametric} and Eq.~\eqref{equation state}. Thus $\kappa_{n}(R,\theta)$ can be converted to $\kappa_{n}(t, h)$. At a negative $h$, the temperature dependence of second- to fourth-order cumulants all have a positive peak $\kappa_{n}^{max}(t,h)$ $(n=2, 3, 4)$. The cumulants can be normalized by their maximum at the peak as follows,
\begin{equation}\label{normalized cumulants}
\kappa_{n}^{Norm}(t,h)=\kappa_{n}(t, h)/\kappa_{n}^{max}(t,h).
\end{equation}

The fixed point behavior has been found in the temperature dependence of $\kappa_{n}^{Norm}(t, h)$ at different $h$ in the three-dimensional Ising model. In order to map the results to that of the QCD, the following linear mapping relations are assumed,
\begin{equation}\label{linear mapping}
\begin{split}
&\frac{T-T_{c}}{\Delta T}=\sin\alpha_1\frac{t}{\Delta t}+\sin\alpha_2 \frac{h}{\Delta h},\\
&\frac{\mu_B-\mu_{Bc}}{\Delta \mu_B}=-\cos\alpha_1\frac{t}{\Delta t}-\cos\alpha_2 \frac{h}{\Delta h}.\\
\end{split}
\end{equation}

Where $T_{c}$, $\mu_{Bc}$ are the temperature and net-baryon chemical potential at the QCD critical point. $\Delta T$ and $\Delta \mu_B$ denote the width of the critical regime in the QCD $T-\mu_B$ plane. Because the location of the critical point and the width of the critical regime for QCD are not known, the suggestion that $\Delta \mu_B \approx 0.1$ GeV from model calculations~\cite{Phys.Rev.D.67.014028} and lattice QCD calculations~\cite{Phys.Rev.D.78.114503} is used. We set $\Delta \mu_B = 0.1$ GeV and $\mu_{Bc} = 0.25$ GeV as was done in Ref.~\cite{Phys.Rev.C.92.034912}. $T_{c}$ is set as $0.165$ GeV and $\Delta T/T_{c} = 1/8$.

$\Delta t$ and $\Delta h$ denote the width of the critical regime in the Ising $t-h$ plane. For details of defining the critical region, see Ref.~\cite{Phys.Rev.C.92.034912}.
After mapping $t$ and $h$ to the QCD $T-\mu_B$ phase diagram, $\alpha_1$ is the angel between the $t$-direction and the opposite direction of $\mu_B$ axis, $\alpha_2$ is the angel between the $h$-direction and the opposite direction of $\mu_B$ axis. A simply sketch of the mapping is shown in Fig.~1(a).

\begin{figure*}[hbt]
\centering
    \includegraphics[width=0.45\textwidth]{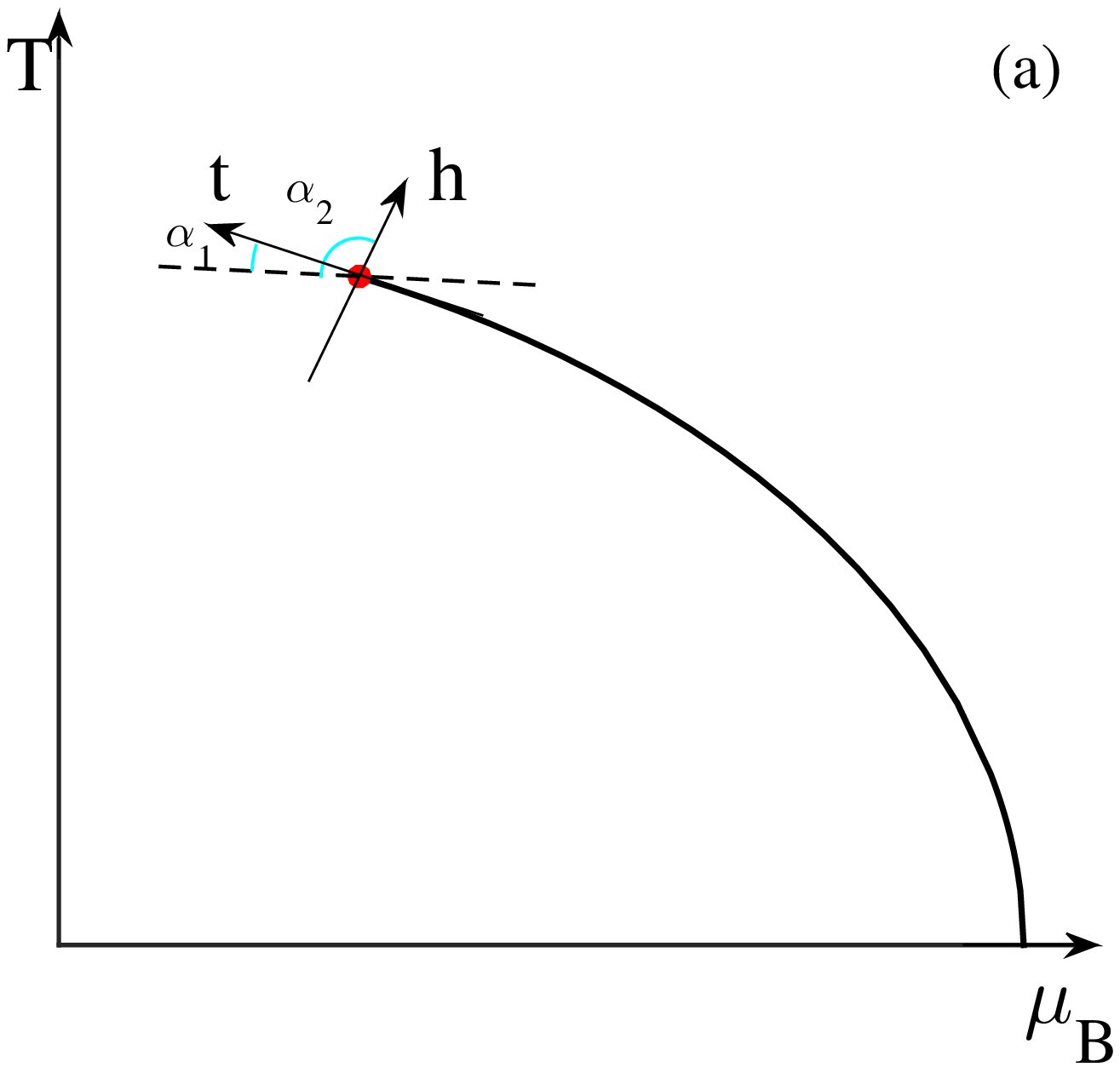}
    \includegraphics[width=0.43\textwidth]{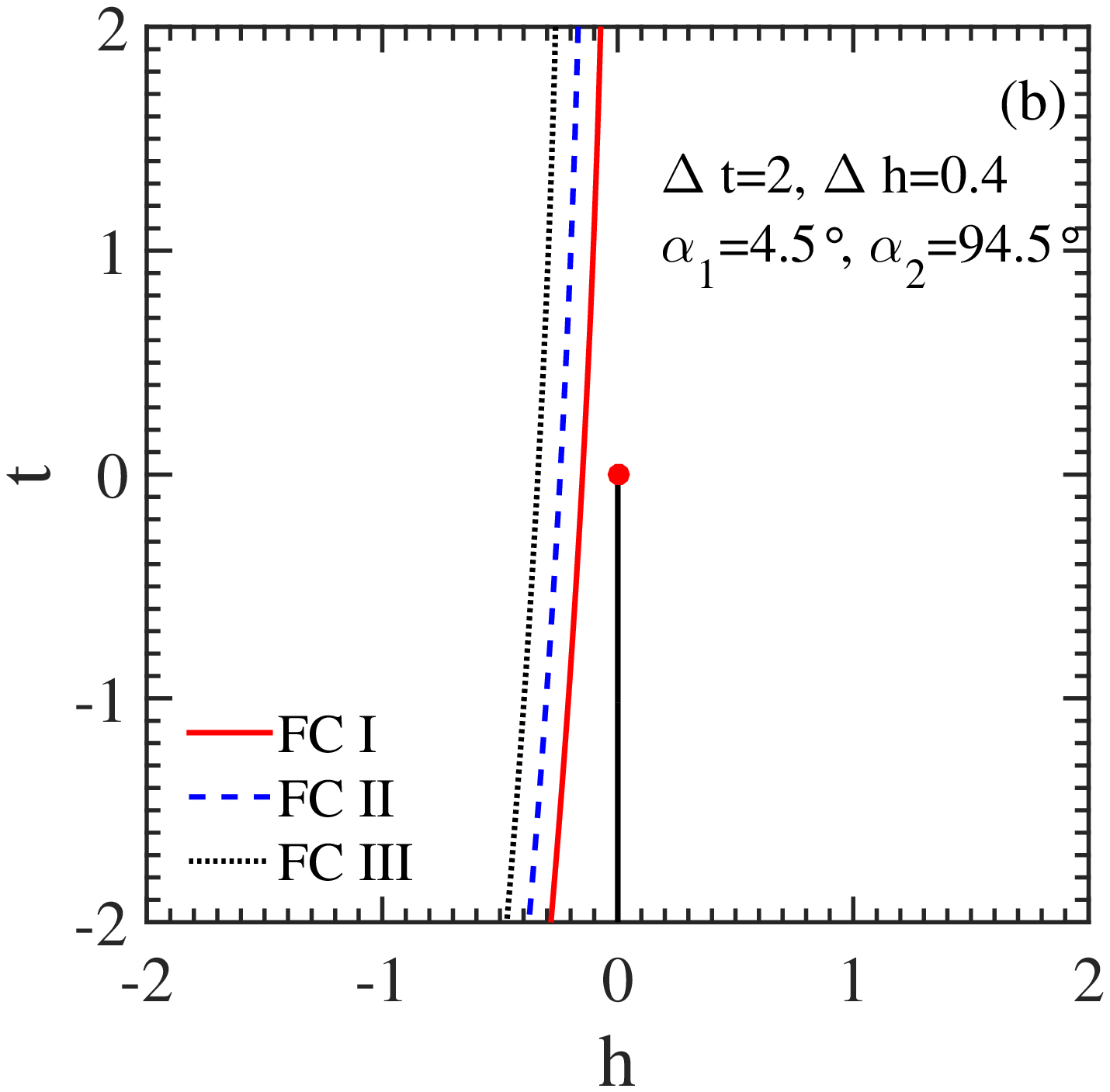}
    \caption{\label{Fig. 1}(Color online). A sketch of the mapping between QCD and Ising variables described by Eq.~\eqref{linear mapping} (a), the mapping of three different freeze-out curves, FC I, FC II and FC III, given by Eqs.~\eqref{freeze-out curve}, \eqref{freeze-out curve I} and \eqref{freeze-out curve II} to the Ising $t-h$ plane by Eq.~\eqref{linear mapping} for $\Delta t=2$, $\Delta h=0.4$, $\alpha_1=4.5^\circ$ and $\alpha_2=94.5^\circ$ (b), respectively.}
\end{figure*}

The freeze-out curve is assumed below the crossover/first-order phase transition line. An empirical parametrization of the heavy-ion-collision data from Ref.~\cite{Phys.Rev.C.73.034905} can be used to describe the freeze-out curve,
\begin{equation}\label{freeze-out curve}
T_f(\mu_B)=a-b{\mu_B}^2-c{\mu_B}^4.
\end{equation}
Where $a=0.166$ GeV, $b=0.139$ GeV$^{-1}$, $c=0.053$ GeV$^{-3}$.

According to the results in Refs.~\cite{Phys.Rev.C.71.054901,AdvancesinHighEnergyPhysics.2021.6611394}, the chemical freeze-out temperature and net-baryon chemical potential could change as the centrality in the relativistic heavy-ion collisions. Although the temperature does not vary much, the net-baryon chemical potential increases from peripheral to central collisions~\cite{AdvancesinHighEnergyPhysics.2021.6611394}. For simplicity, we move the freeze-out curve described by Eq.~\eqref{freeze-out curve} parallelly to the lower temperature side to get the other two freeze-out curves, i.e.,
\begin{equation}\label{freeze-out curve I}
T_{f1}(\mu_B)=a-b{\mu_B}^2-c{\mu_B}^4-0.005,
\end{equation}
and
\begin{equation}\label{freeze-out curve II}
T_{f2}(\mu_B)=a-b{\mu_B}^2-c{\mu_B}^4-0.01.
\end{equation}

Based on the mapping relation and using Eq.~\eqref{linear mapping}, these three freeze-out curves can be converted to three lines on the Ising $t-h$ plane. As an example, supposing $\Delta t=2$, $\Delta h=0.4$, the angels $\alpha_1=4.5^{\circ}$ and $\alpha_2=94.5^{\circ}$, a sketch of the mapping is shown in Fig.~1(b). The solid red, dashed blue and dotted black line from right to left represent the mapping of the freeze-out curves FC I, FC II and FC III presented by Eqs.~\eqref{freeze-out curve}, \eqref{freeze-out curve I} and \eqref{freeze-out curve II}, respectively. It is clear that for each freeze-out curve after mapping to the Ising $t-h$ plane, the external magnetic field is not fixed as the cases in Ref.~\cite{cpc2022}, but obvious fixed point behavior still can be observed in the temperature dependence of $\kappa_2^{Norm}$, $\kappa_3^{Norm}$ and $\kappa_4^{Norm}$ along these three freeze-out curves, which are presented in Fig.~2.

\begin{figure*}[hbt]
\centering
    \includegraphics[width=0.32\textwidth]{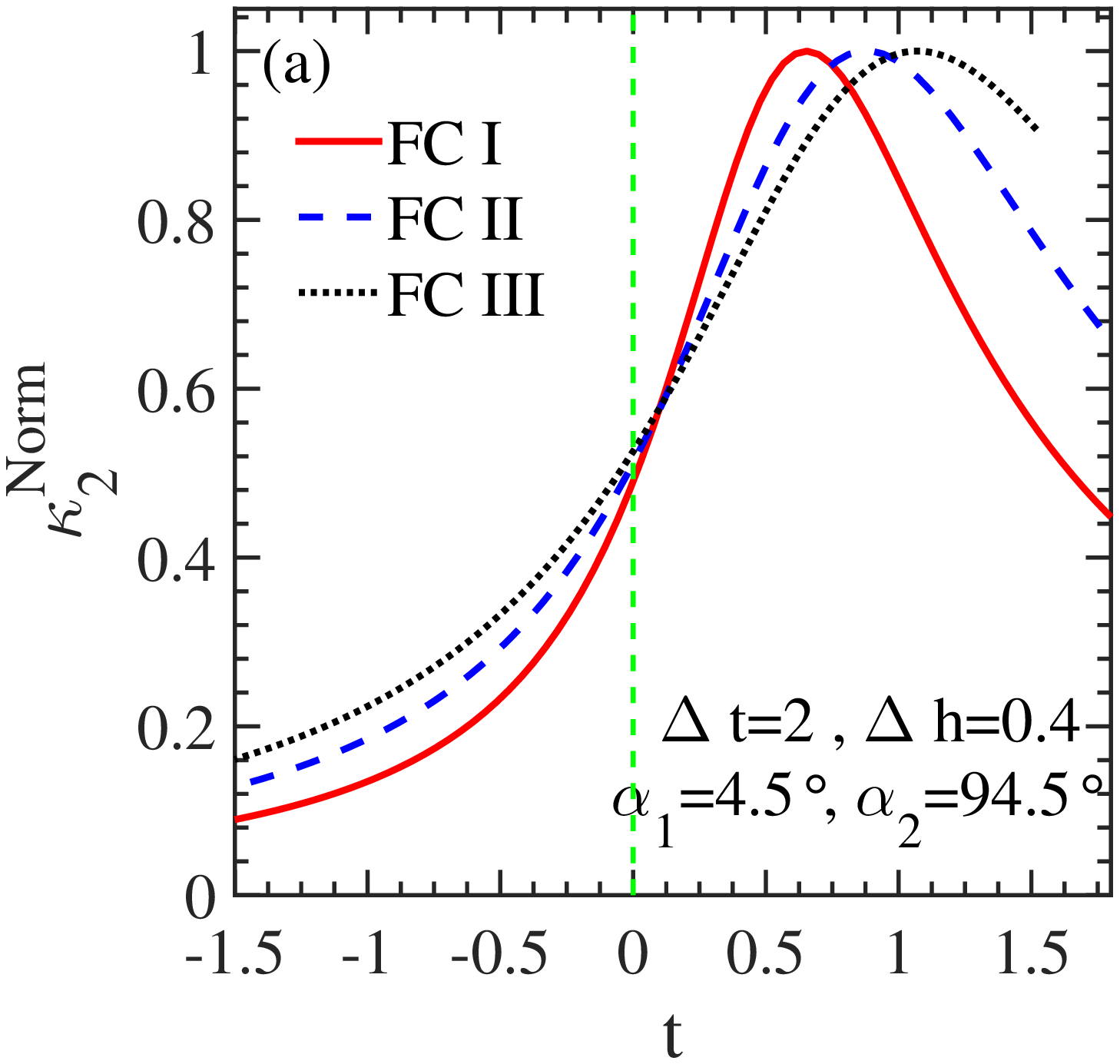}
    \includegraphics[width=0.32\textwidth]{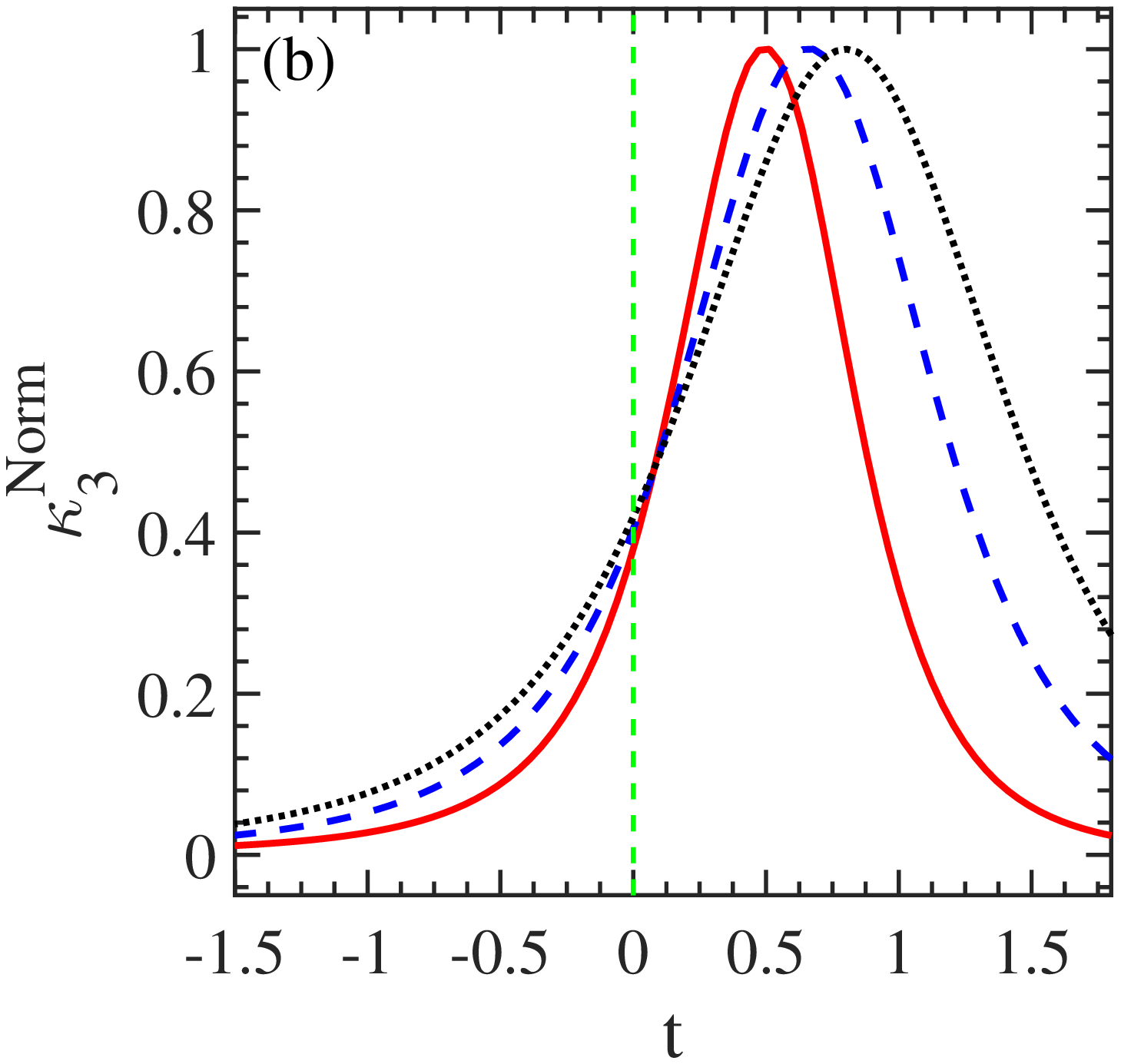}
    \includegraphics[width=0.32\textwidth]{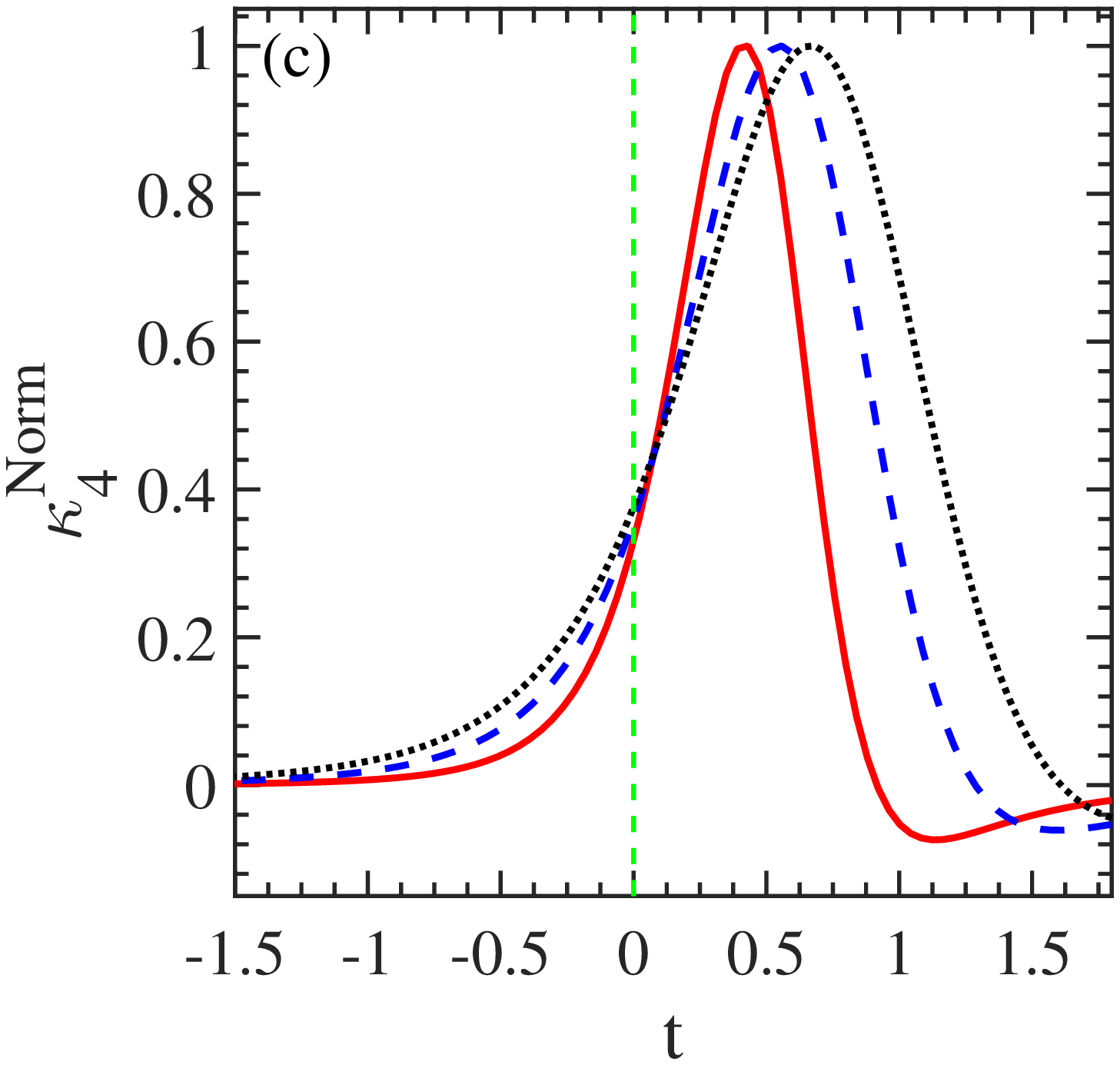}
    \caption{\label{Fig. 2}(Color online). Temperature dependence of $\kappa_2^{Norm}$,  $\kappa_3^{Norm}$, and $\kappa_4^{Norm}$ for $\Delta t = 2$, $\Delta h = 0.4$, $\alpha_1 = 4.5^{\circ}$ and $\alpha_2 = 94.5^{\circ}$ along the three different freeze-out curves, respectively. The vertical green dashed line shows the critical temperature $t=0$.}
\end{figure*}

It is clear that the fixed point behavior exists in each sub-figure of Fig.~2. Although the temperature at the fixed point is not the critical temperature and a little bigger than the critical one, i.e. the fixed point is at the right side of the vertical green dashed line $t=0$, it is still in the vicinity of the critical temperature. What is more, the fixed point is closer to the vertical green dashed line than the peak in the cumulants.

\section{Fixed point behavior in $\mu_B$ dependence of the normalized cumulants}

Through the linear mapping relation between QCD and Ising variables represented by Eq.~\eqref{linear mapping}, $t$ dependence of the cumulants in the Ising model can be converted to $\mu_B$ dependence. There are four parameters $\alpha_1$, $\alpha_2$, $\Delta t$ and $\Delta h$. Along the $t$-direction, the symmetry is preserved. It is easy to realize that the $t$-direction should be tangential to the QCD first-order phase transition line at the critical point. As a result, the angel $\alpha_1$ is usually not set very large~\cite{Phys.Rev.C.103.034901}. Here it is fixed at $4.5^\circ$, which is close to the value in Ref.~\cite{Phys.Rev.C.103.034901}, where it is set as $4.6^\circ$. For the $h$-direction after mapping to QCD, there is no general rule. To set the $h$-direction orthogonal to the $t$-direction is the common and simply choice in the literature. That is $\alpha_2=94.5^\circ$ here. To study the influence of $h$-direction slightly deviated from the orthogonal direction of $t$, the other two values are also chosen for $\alpha_2$. They are $72^\circ$ and $117^\circ$, respectively. What is more, $\alpha_2$ and $\alpha_2 \pm \pi$ are equivalent for the mapping since the sign of $h$ is a matter of convention.

After the values of $\Delta t$ and $\Delta h$ are set as $2$ and $0.4$, respectively, the influence of $\alpha_2$ on the fixed point behavior in the $\mu_B$ dependence of the normalized second- to fourth-order cumulants are studied and shown in Fig.~3.

\begin{figure*}[hbt]
\centering
    \includegraphics[width=0.31\textwidth]{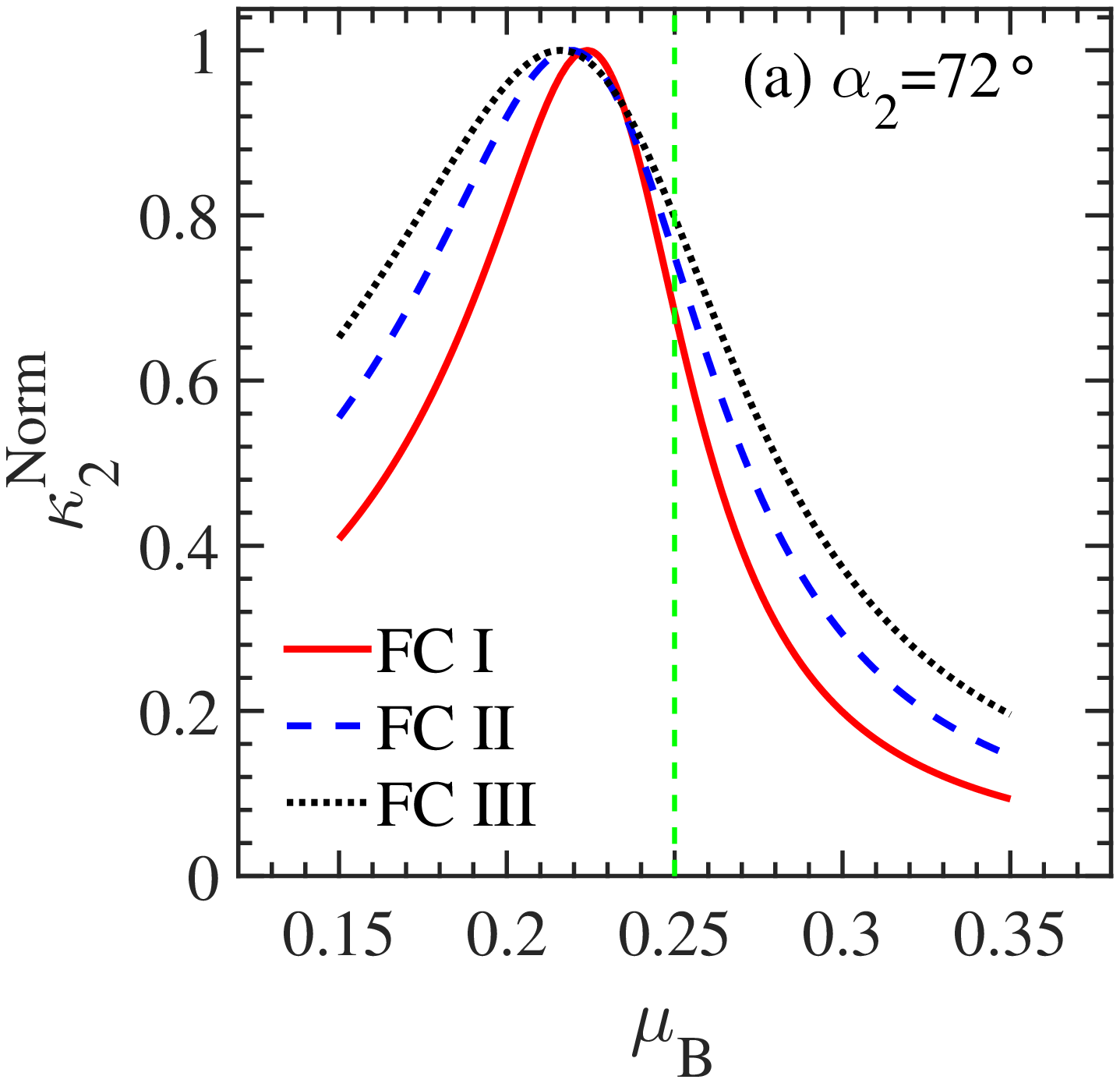}
    \includegraphics[width=0.31\textwidth]{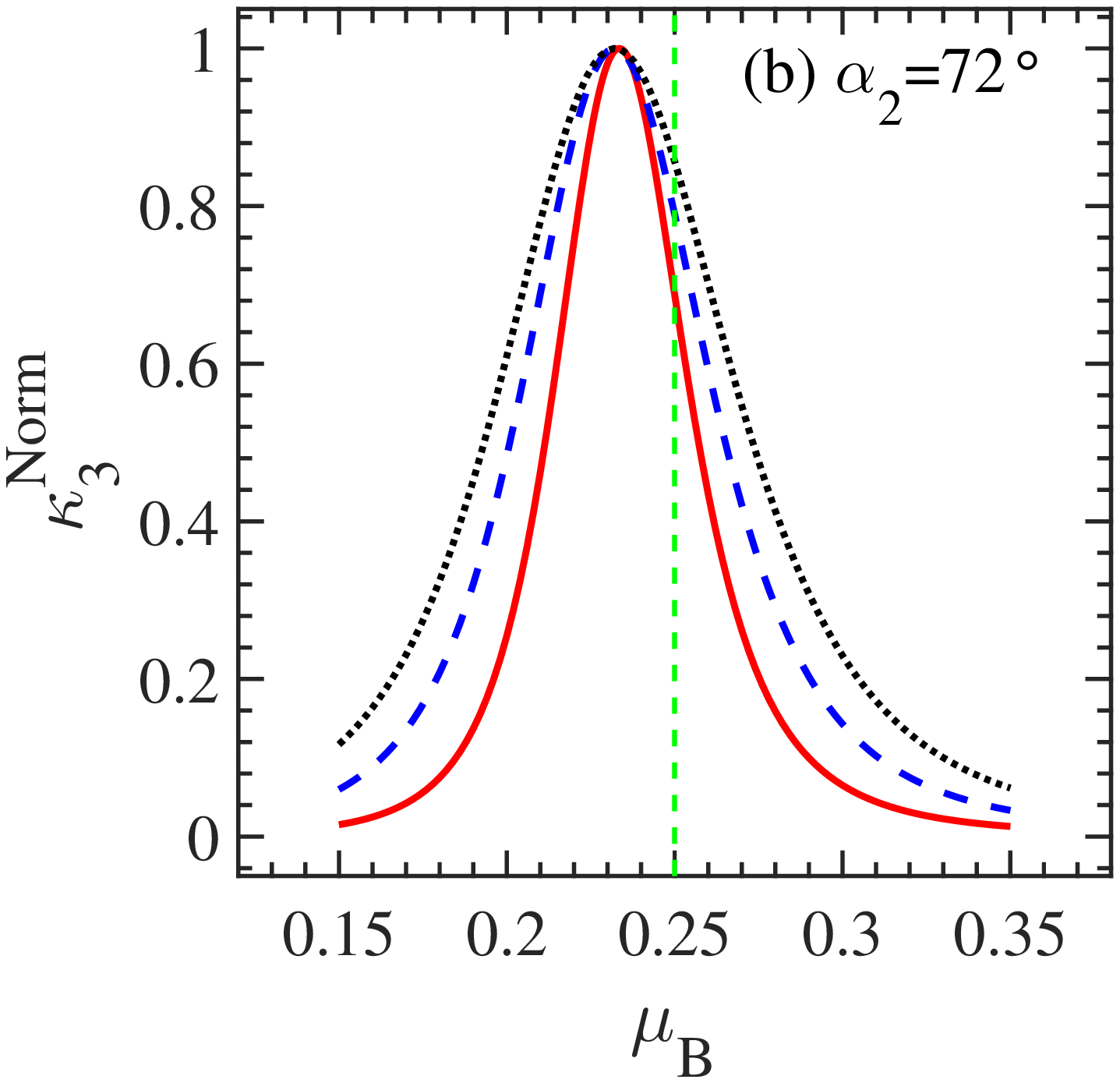}
    \includegraphics[width=0.31\textwidth]{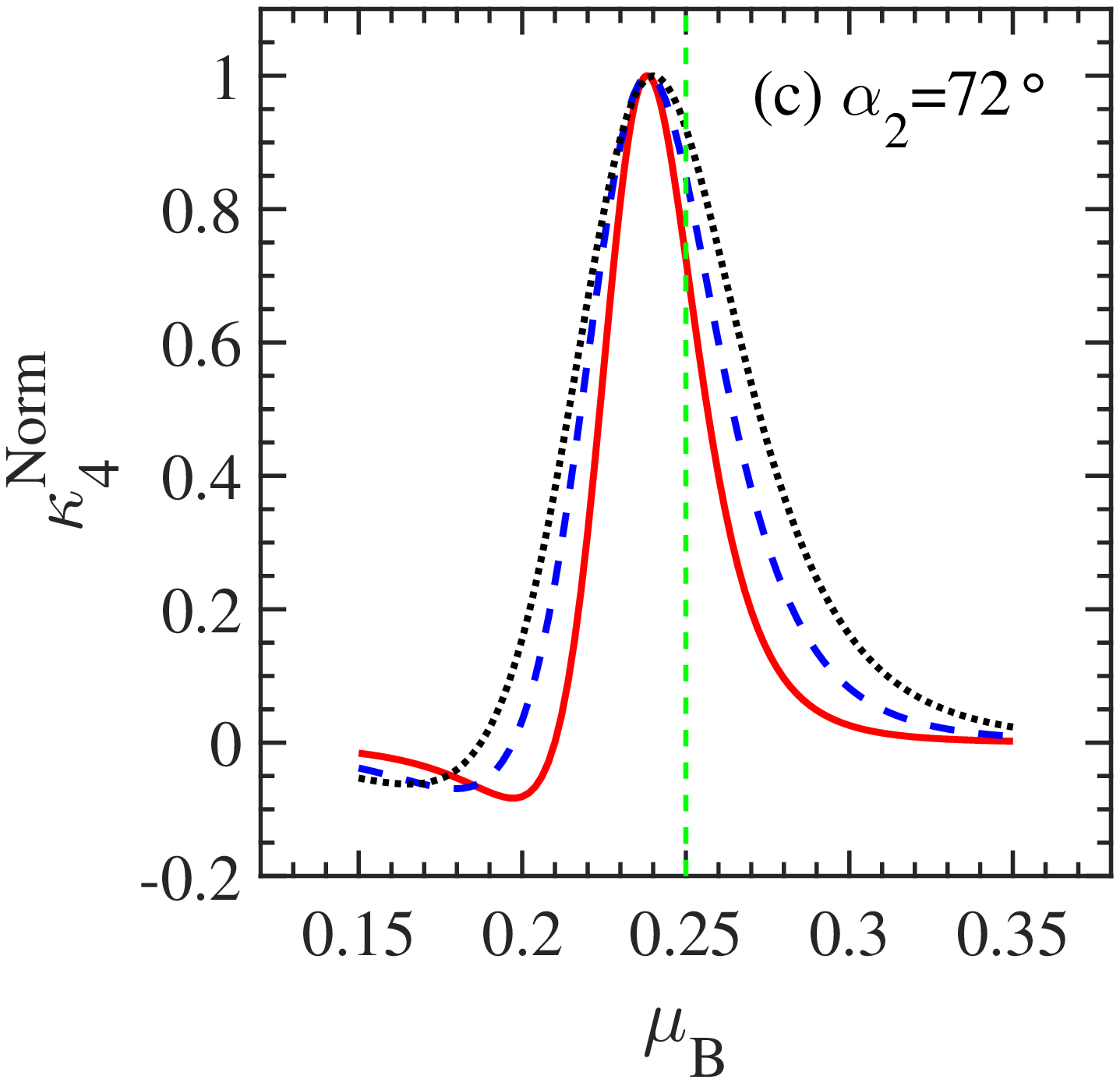}
    \includegraphics[width=0.31\textwidth]{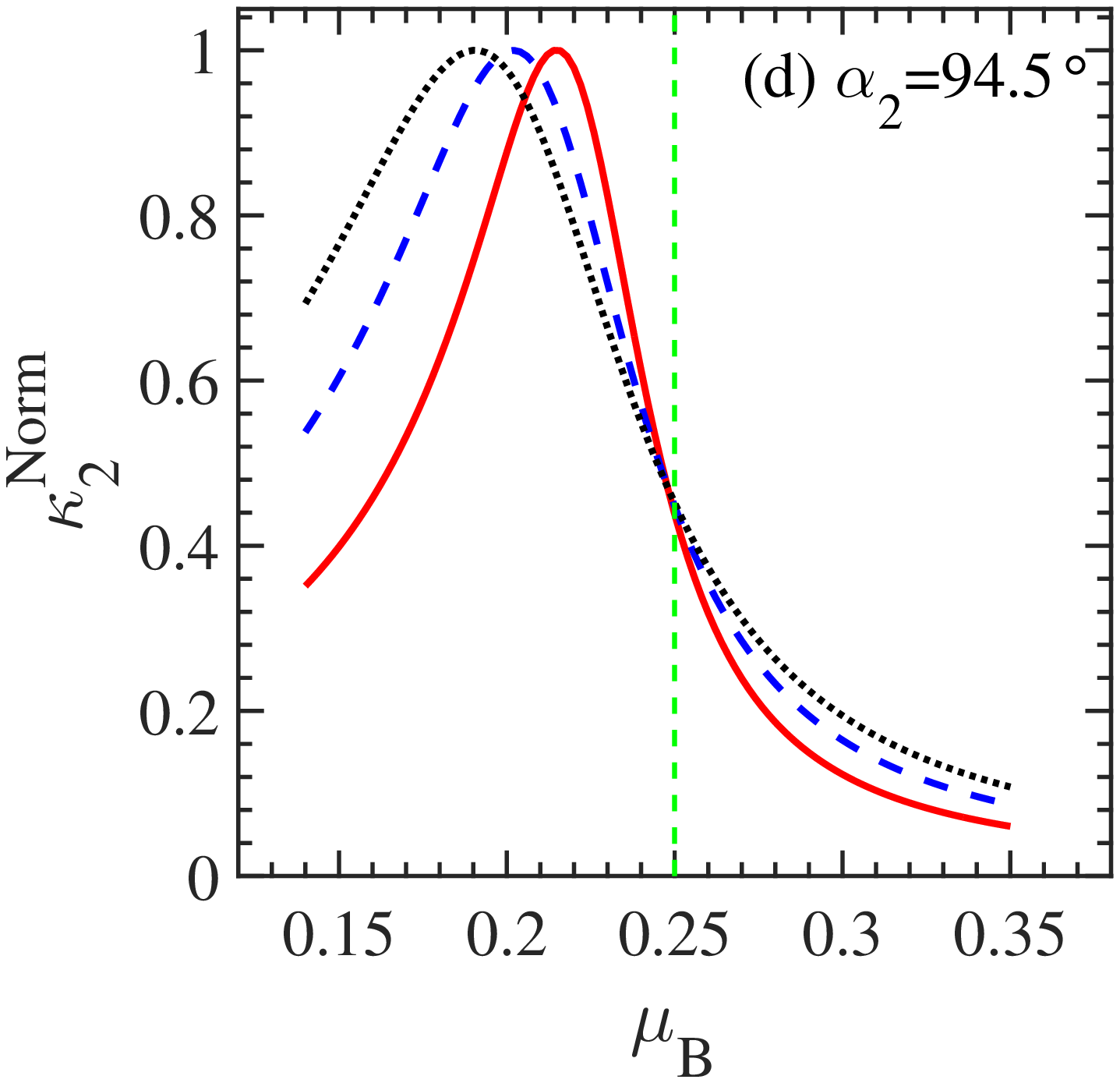}
    \includegraphics[width=0.31\textwidth]{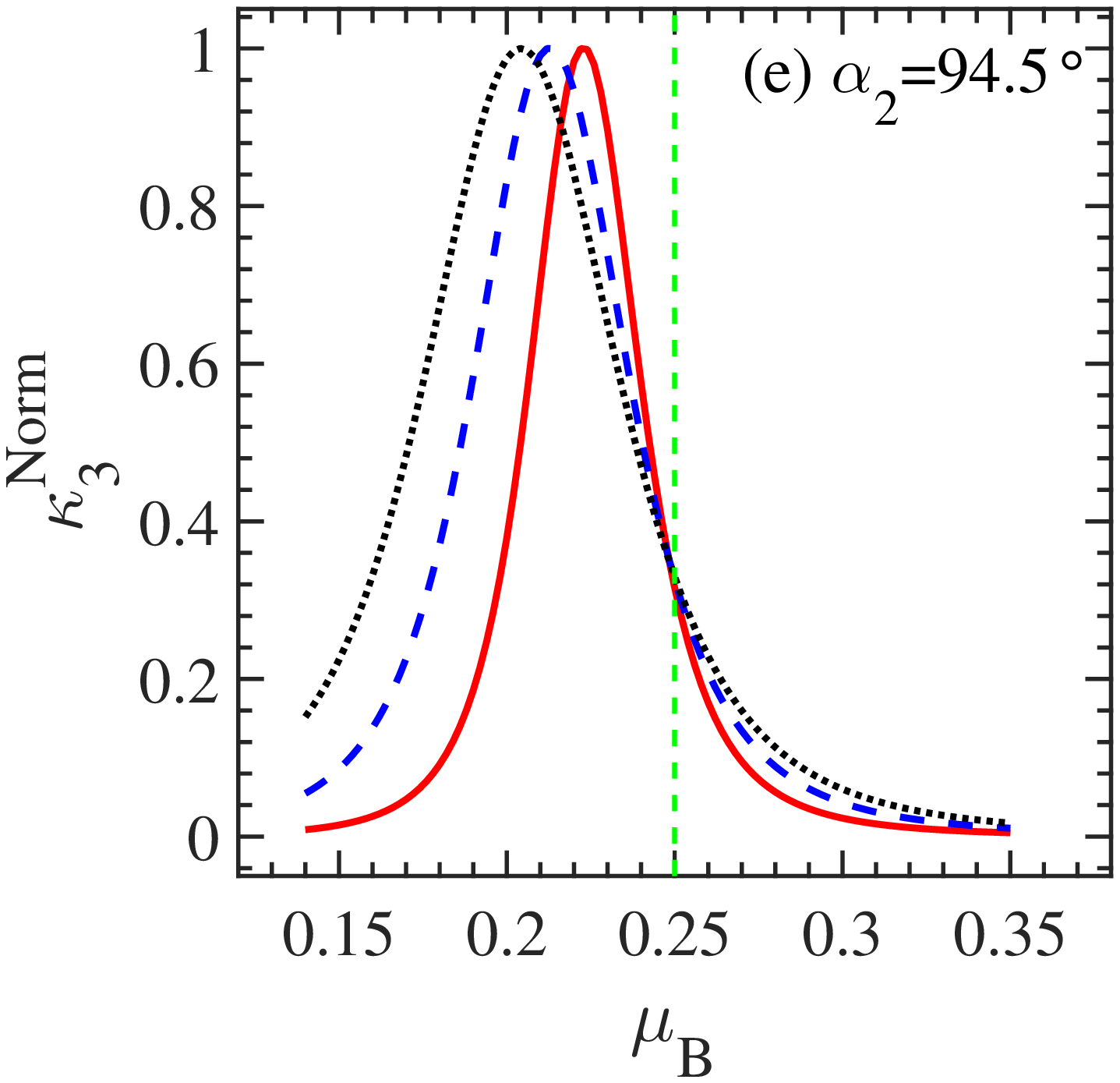}
    \includegraphics[width=0.31\textwidth]{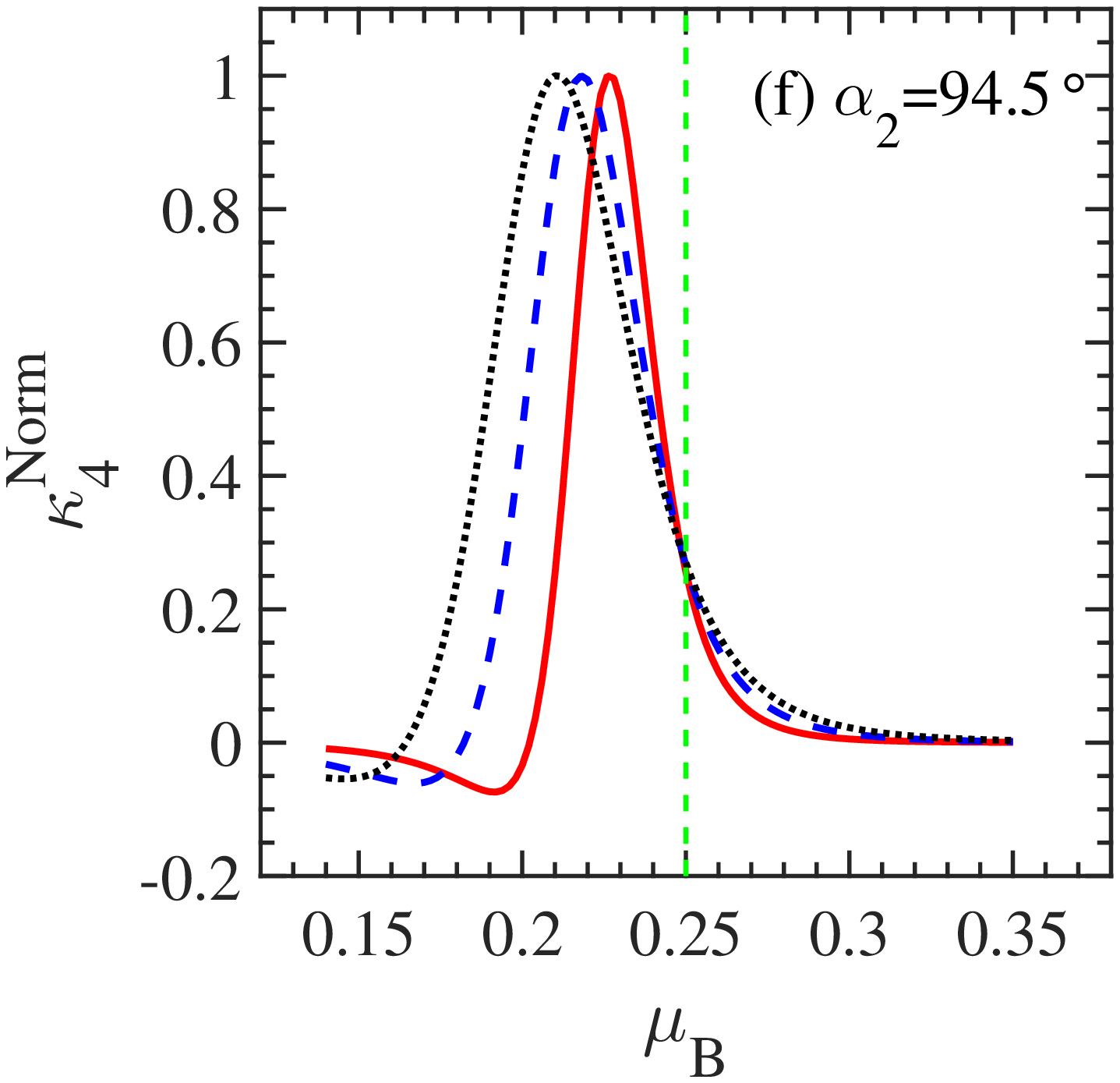}
    \includegraphics[width=0.31\textwidth]{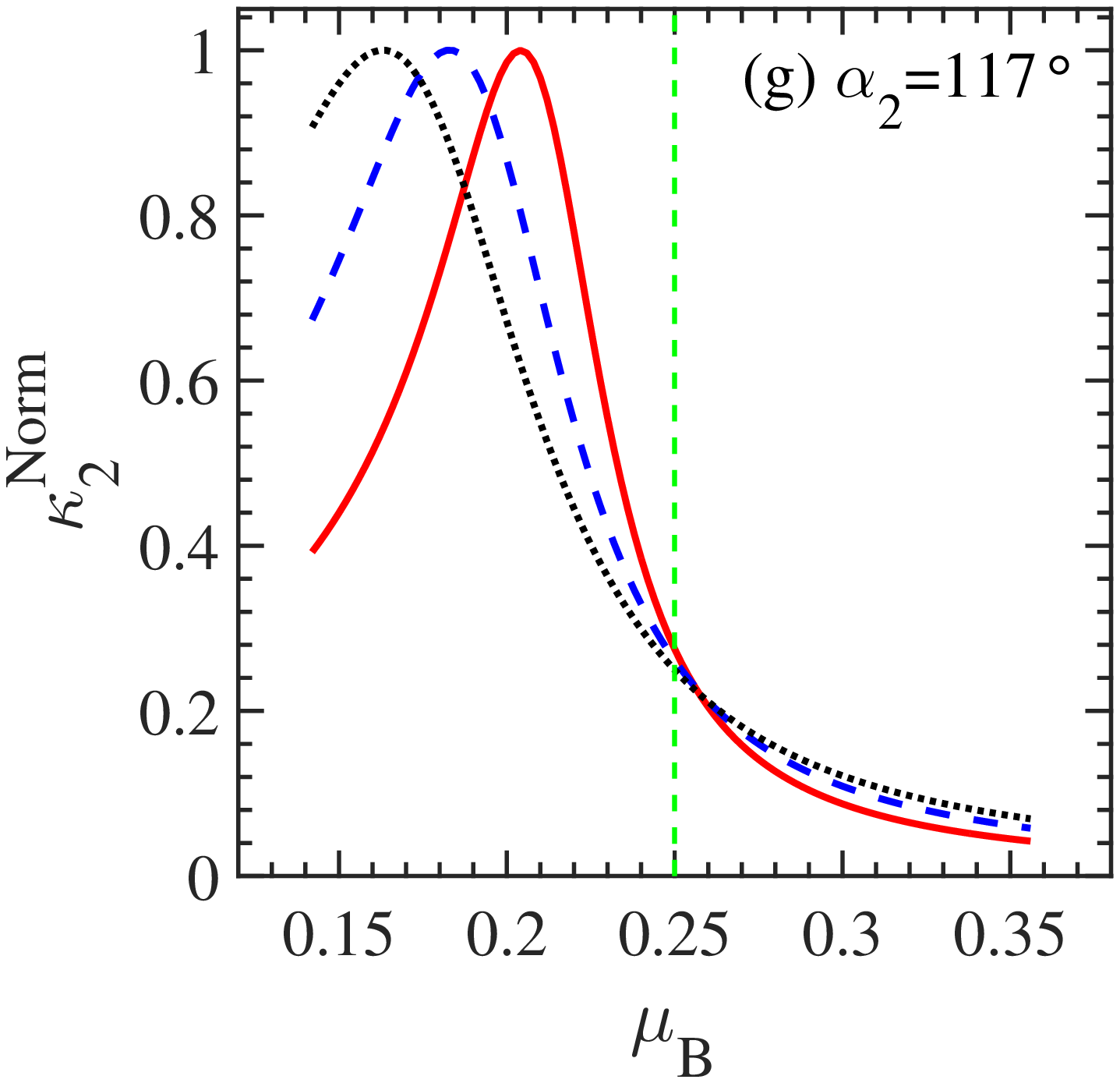}
    \includegraphics[width=0.31\textwidth]{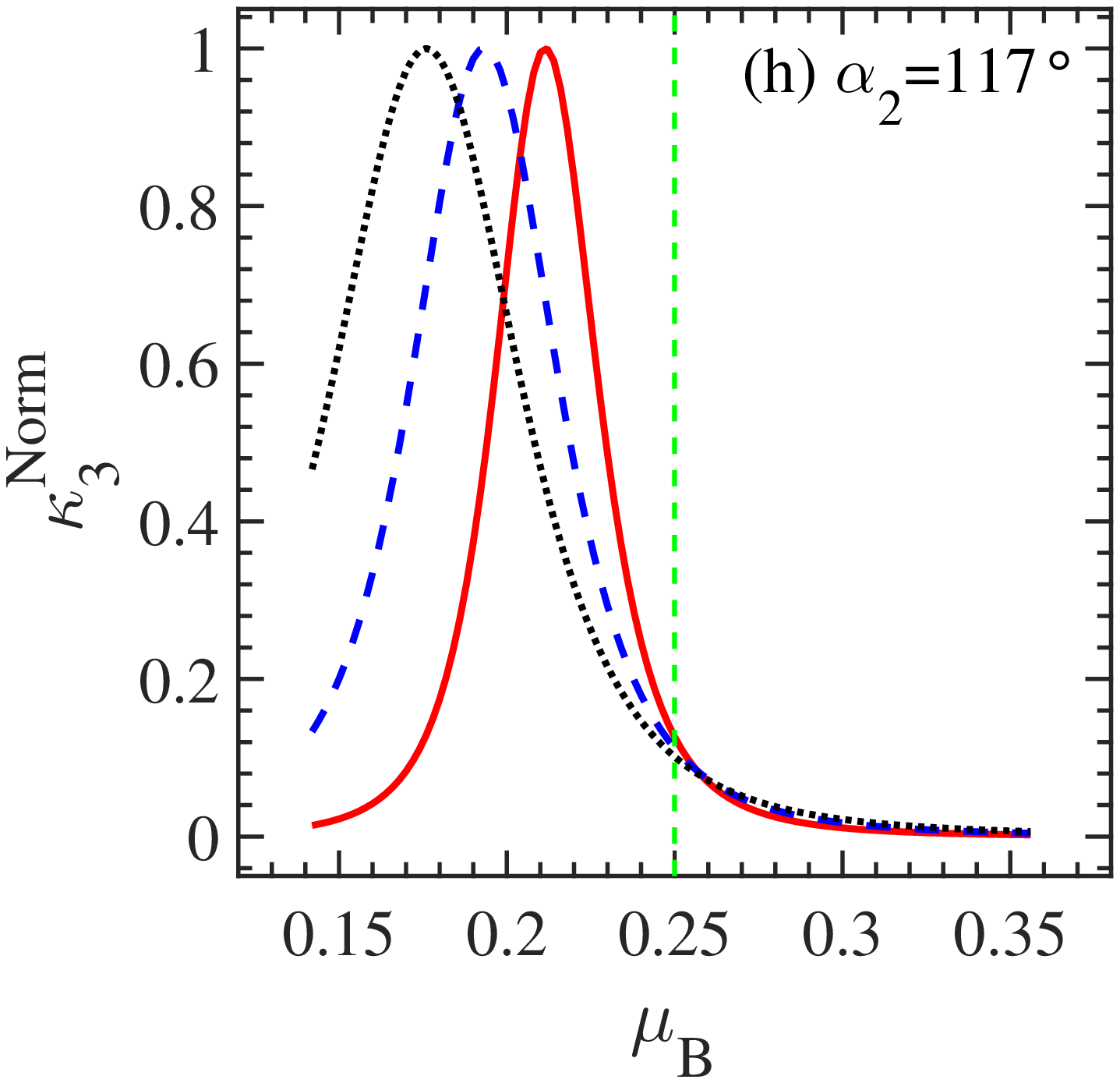}
    \includegraphics[width=0.31\textwidth]{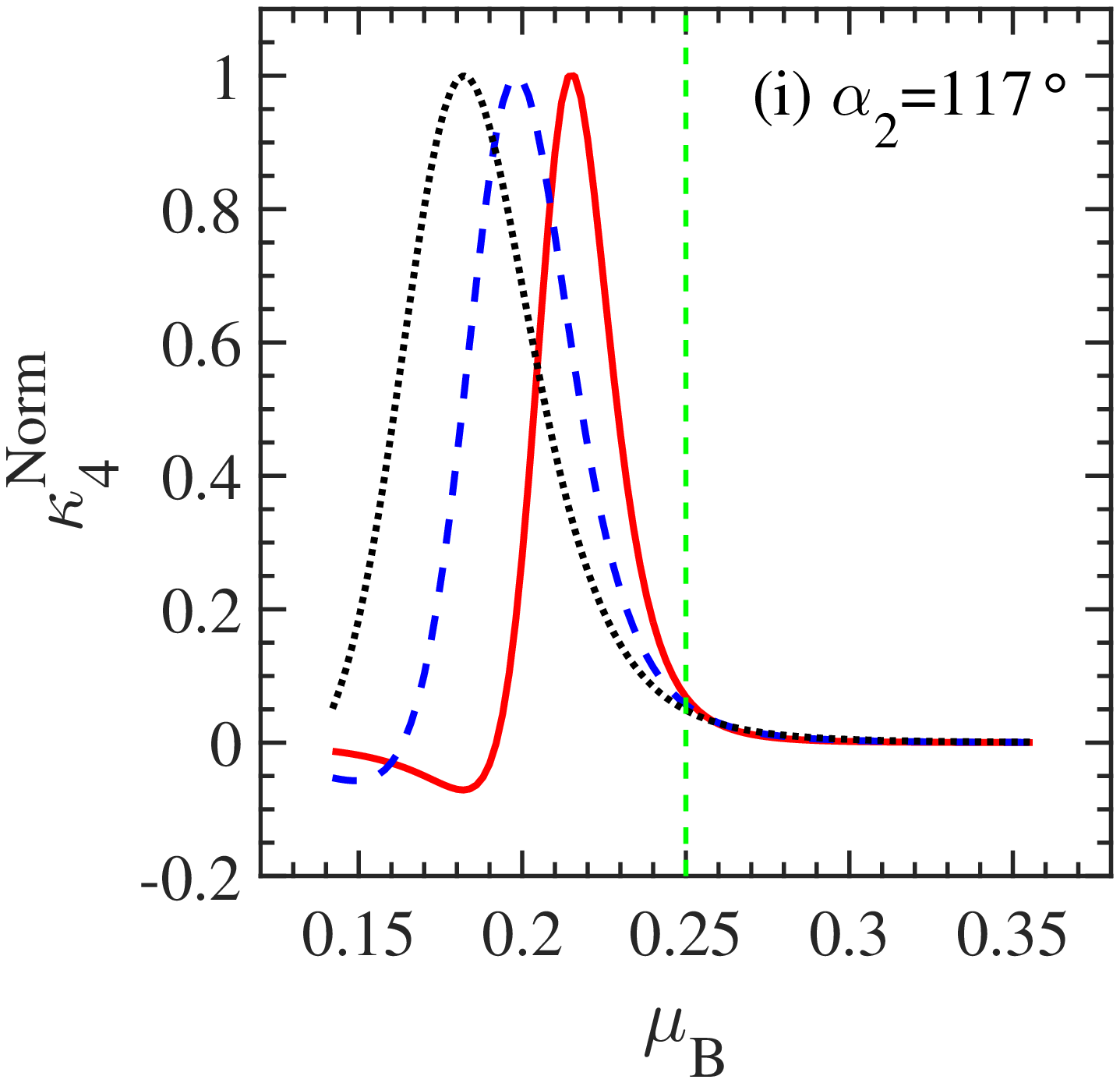}
    \caption{\label{Fig. 3}(Color online). Net-baryon chemical potential dependence of $\kappa_2^{Norm}$,  $\kappa_3^{Norm}$, and $\kappa_4^{Norm}$ at $\alpha_2 = 72^{\circ}$ (up panels), $94.5^{\circ}$ (middle panels) and $117^{\circ}$ (down panels) with $\alpha_1 = 4.5^{\circ}$, $\Delta t = 2$ and $\Delta h = 0.4$ along three different freeze-out curves, respectively. The green dashed line shows the net-baryon chemical potential at the QCD critical point.}
\end{figure*}

In each sub-figure of Fig.~3, the vertical dashed green line $\mu_B=\mu_{Bc}$ shows the critical net-baryon chemical potential. The solid red curve, dashed blue curve and dotted black curve represent the $\mu_B$ dependence of the normalized cumulants along the three freeze-out curves represented by Eq.~\eqref{freeze-out curve}, Eq.~\eqref{freeze-out curve I} and Eq.~\eqref{freeze-out curve II}, respectively. The qualitative $\mu_B$ dependence of each order cumulant does not change along the different freeze-out curves. They all have a peak when approaching $\mu_{Bc}$ from the smaller $\mu_B$ side. The peak gets sharper and closer to the vertical dashed green line when the freeze-out curve gets closer to the phase boundary. I.e. the peak is sharpest in the solid red curve and it is closest to the vertical dashed green line. And the case is just opposite in the dotted black curve.

It is clear that the $\mu_B$ dependence of normalized cumulants along three different freeze-out curves intersect at a point. Except for $\kappa_4^{Norm}$ at $\alpha_2=72^{\circ}$ in Fig.~3(c), the fixed points are all at the right side of the peaks and closer to the vertical dashed green line than the peaks.

With the three different values of $\alpha_2$ chosen here, the fixed point behavior all exits, while the position of the fixed point moves a little. In the top row of Fig.~3, $\alpha_2=72^\circ$, the fixed point is at the left side of the vertical dashed green line. $\mu_B$ of the fixed point is smaller than $\mu_{Bc}$. With $\alpha_2=94.5^\circ$, i.e. the middle row of Fig.~3, the fixed point is nearly on the vertical dashed green line. Its $\mu_B$ is very close to $\mu_{Bc}$. While with $\alpha_2=117^\circ$, i.e. the bottom row of Fig.~3, the fixed point is at the right side of the vertical dashed green line. Its $\mu_B$ is a little larger than $\mu_{Bc}$. What is more, as the increase of $\alpha_2$, the value of $\kappa_2^{Norm}$, $\kappa_3^{Norm}$ and $\kappa_4^{Norm}$ at the fixed point all decreases. And the higher the order of the normalized cumulants, the faster its value changes with $\alpha_2$.

Although it is not shown here, it should be clarified that if $\alpha_2$ continues to decrease, the fixed point will come to the left side of the peak, just like the case in Fig.~3(c), then the peak will be closer to $\mu_{Bc}$. In reverse, if $\alpha_2$ continues to increase, the value of the normalized cumulant at the fixed point along the three freeze-out curves will approach to zero, just like the case in Fig.3~(i). If $\alpha_2$ approaches to $\alpha_1$ or $\pi \pm \alpha_1$, that is the $t$-direction and $h$-direction are nearly parallel to each other after mapped to the QCD $T-\mu_B$ plane, the three freeze-out curves will not pass through or even reach the critical region of the Ising model. The fixed point behavior in the $t$ dependence of the normalized cumulants along the freeze-out curves does not exist any more.

On the whole, when the $h$-direction and $t$-direction after mapping to the QCD phase diagram is orthogonal or not far from orthogonal, the fixed point behavior exists in $\mu_B$ dependence of normalized cumulants along different freeze-out curves and is closer to $\mu_{Bc}$ than the peak in the cumulants. While the value of $\alpha_2$ will influence the position of the fixed point. When $h$-direction and $t$-direction are nearly orthogonal to each other, $\mu_B$ of the fixed point is closer to $\mu_{Bc}$.

\begin{figure*}[hbt]
\centering
    \includegraphics[width=0.31\textwidth]{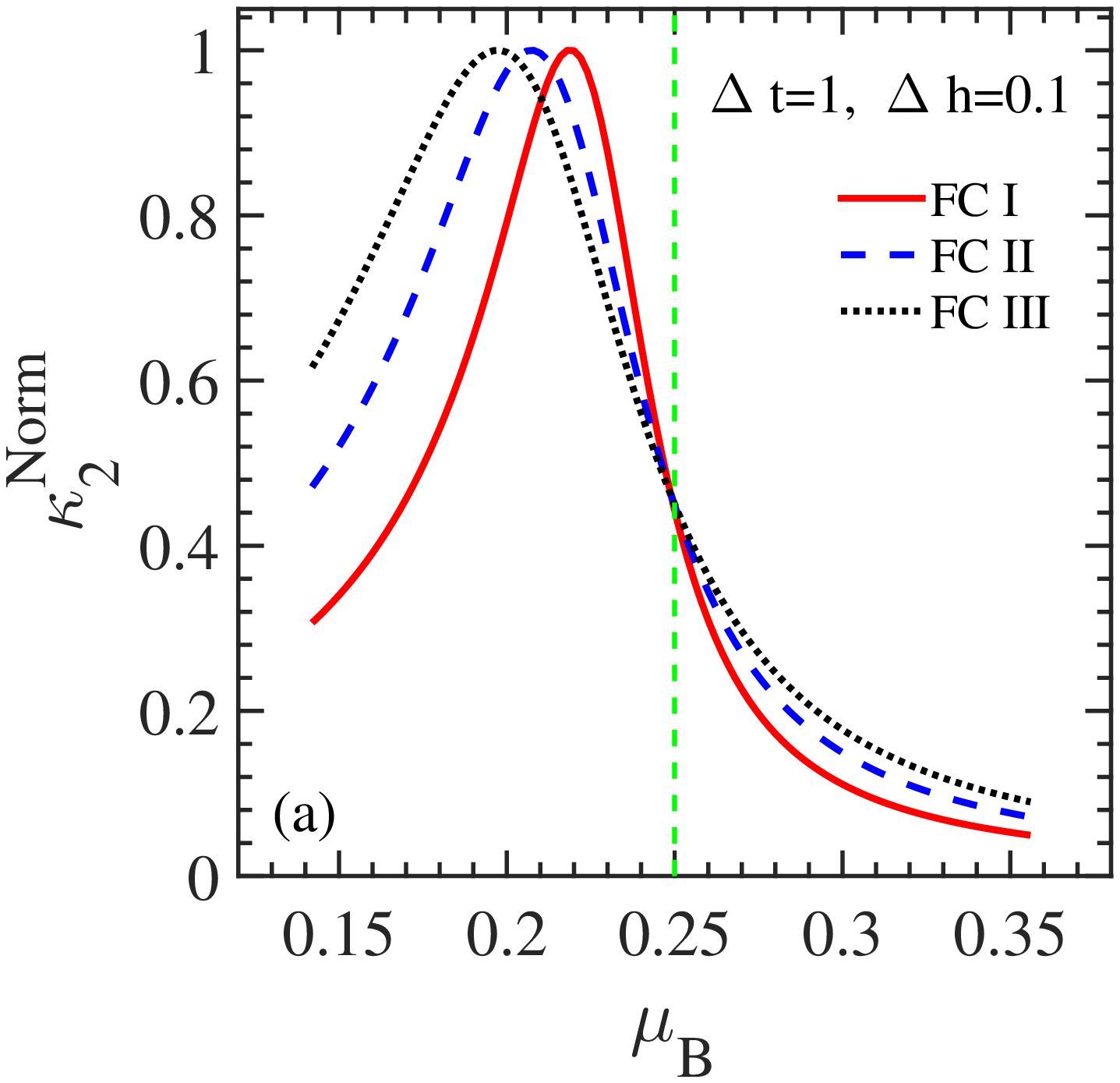}
    \includegraphics[width=0.31\textwidth]{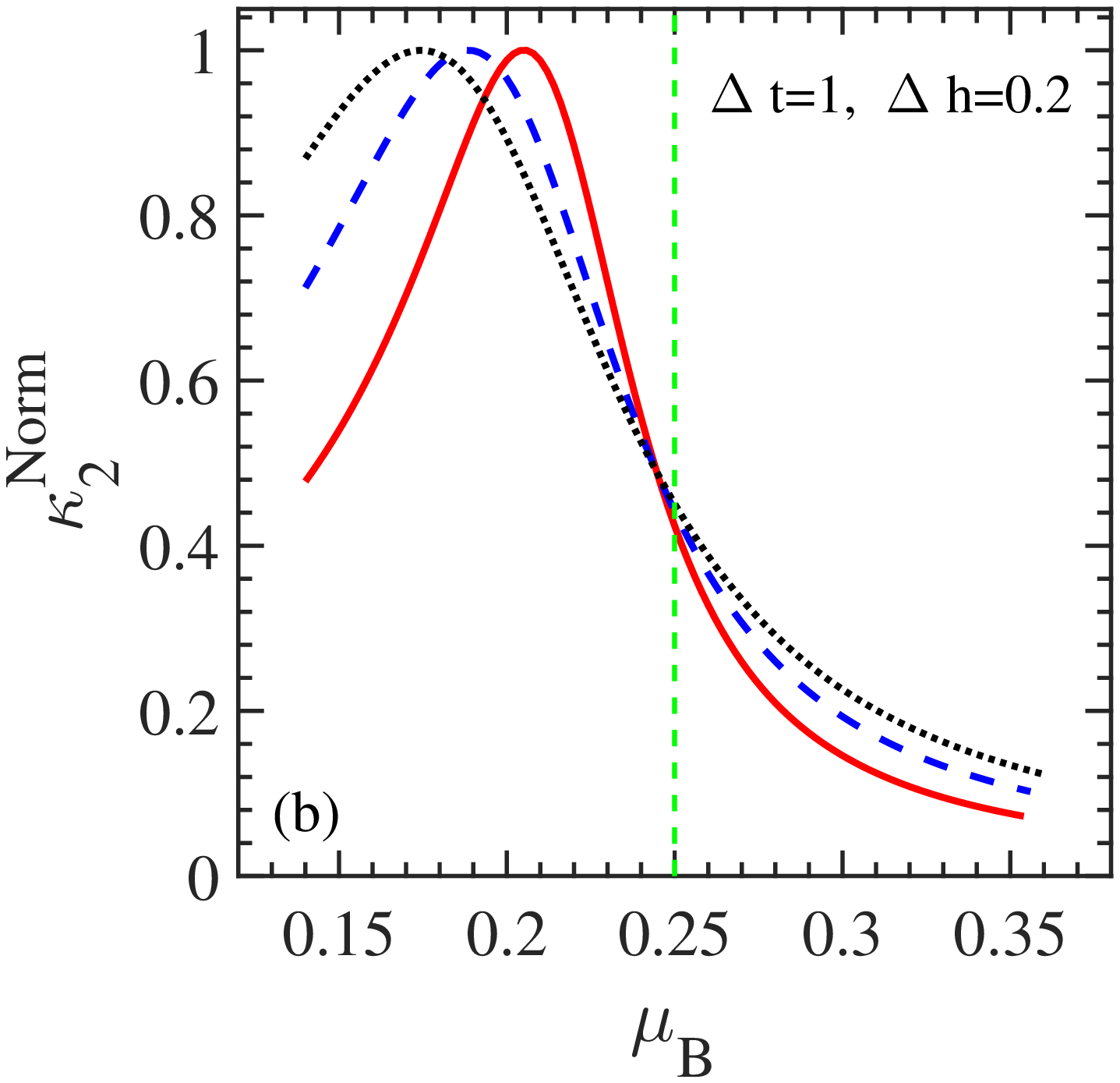}
    \includegraphics[width=0.31\textwidth]{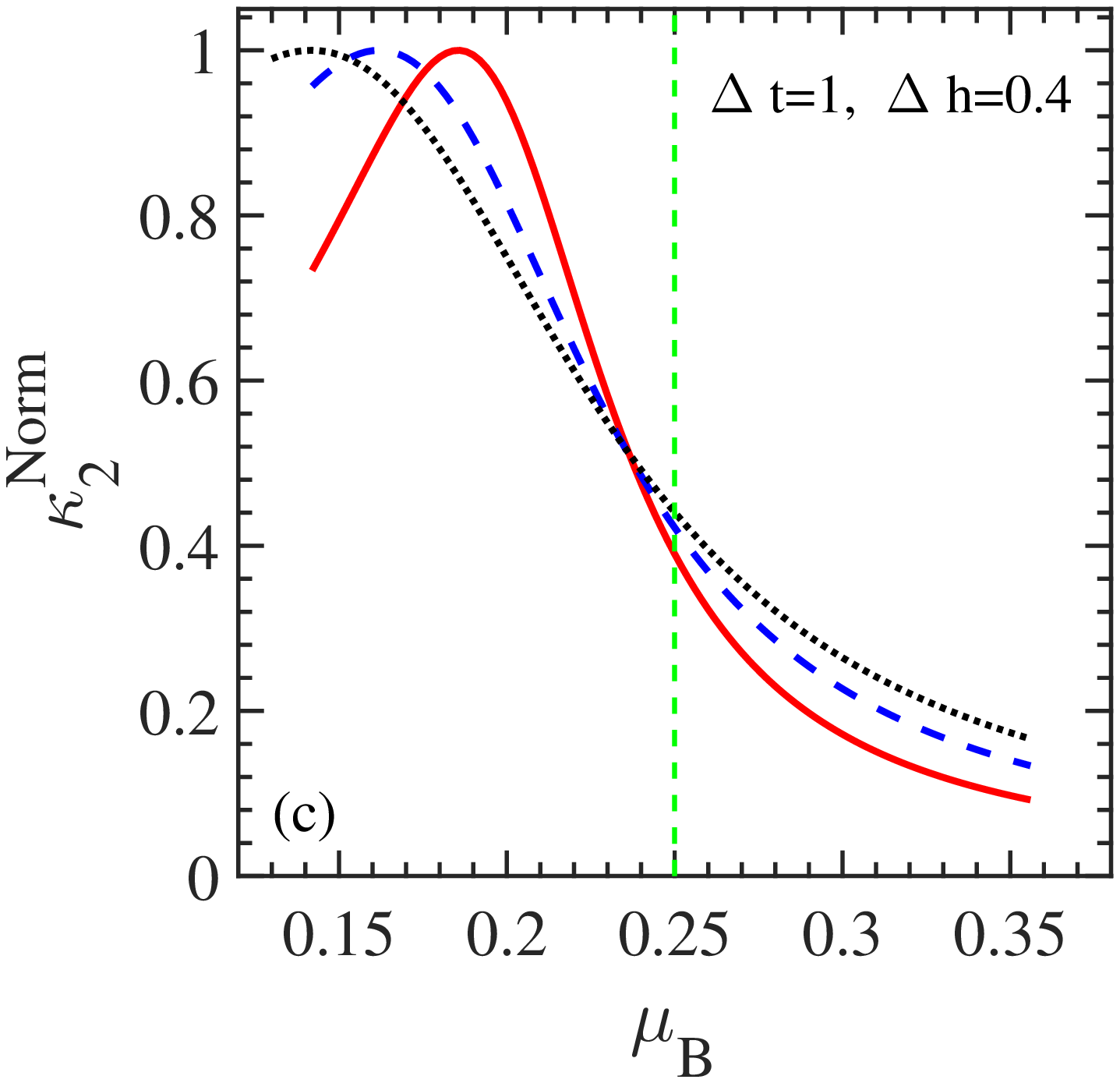}
    \includegraphics[width=0.31\textwidth]{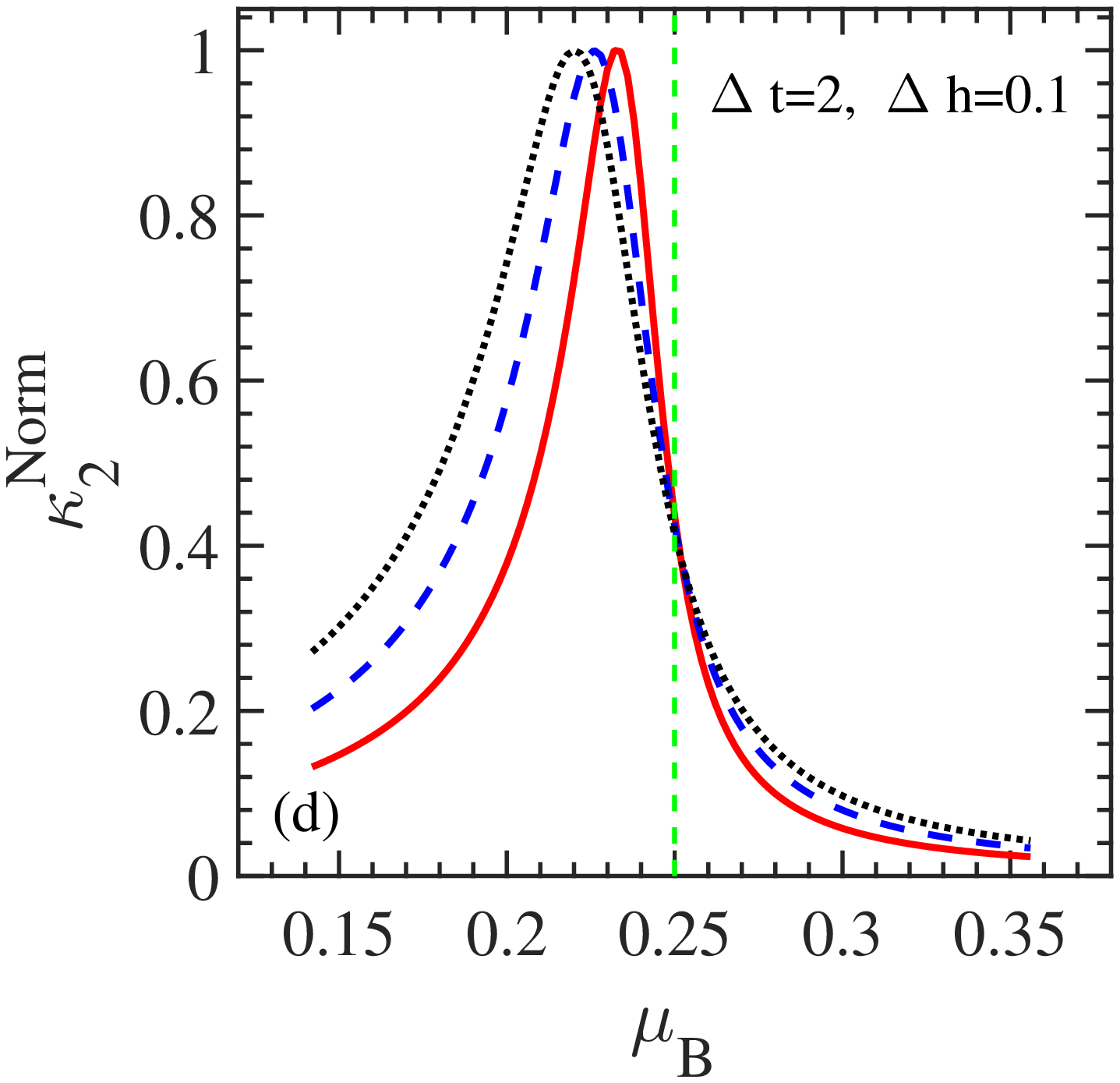}
    \includegraphics[width=0.31\textwidth]{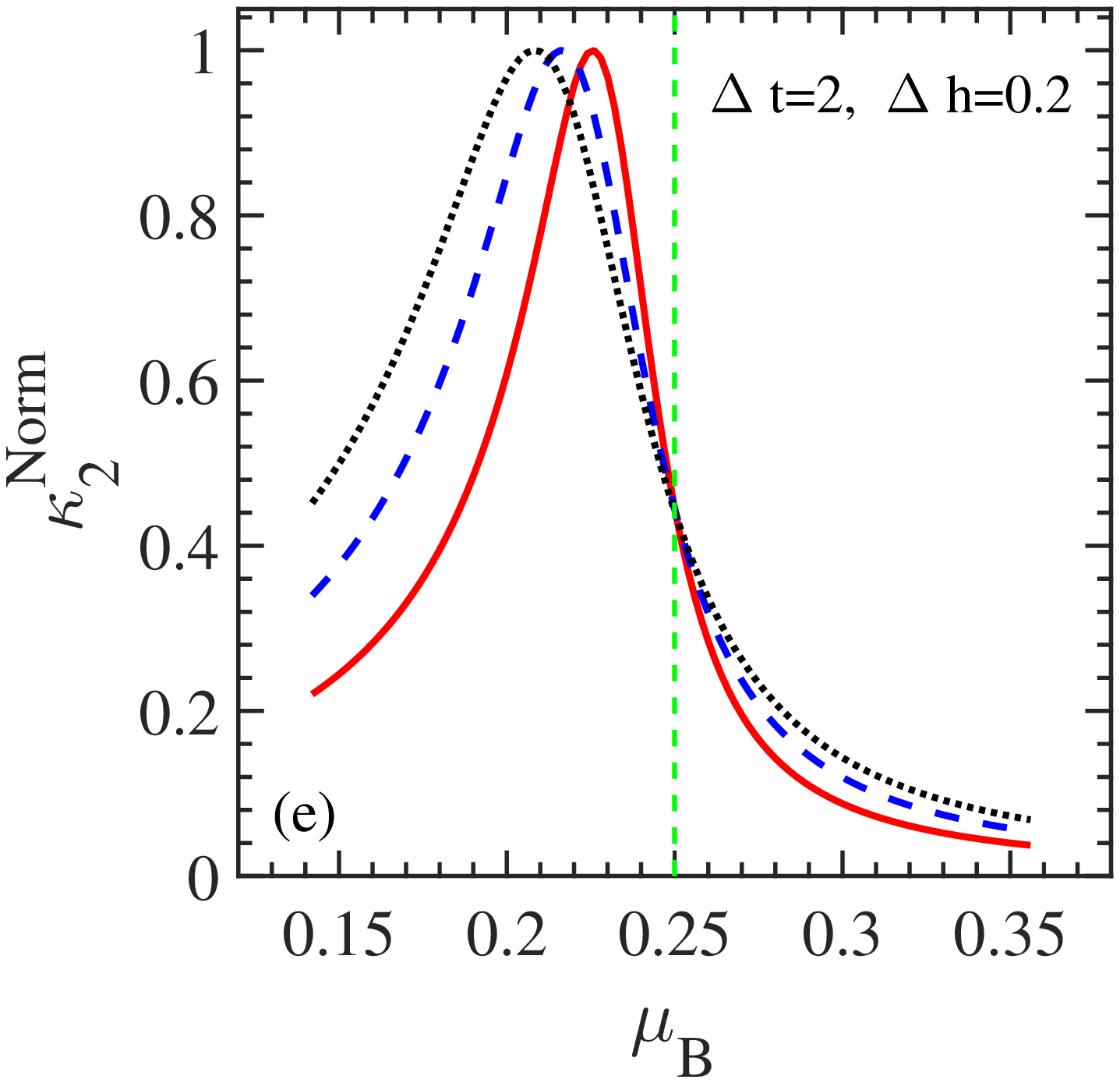}
    \includegraphics[width=0.31\textwidth]{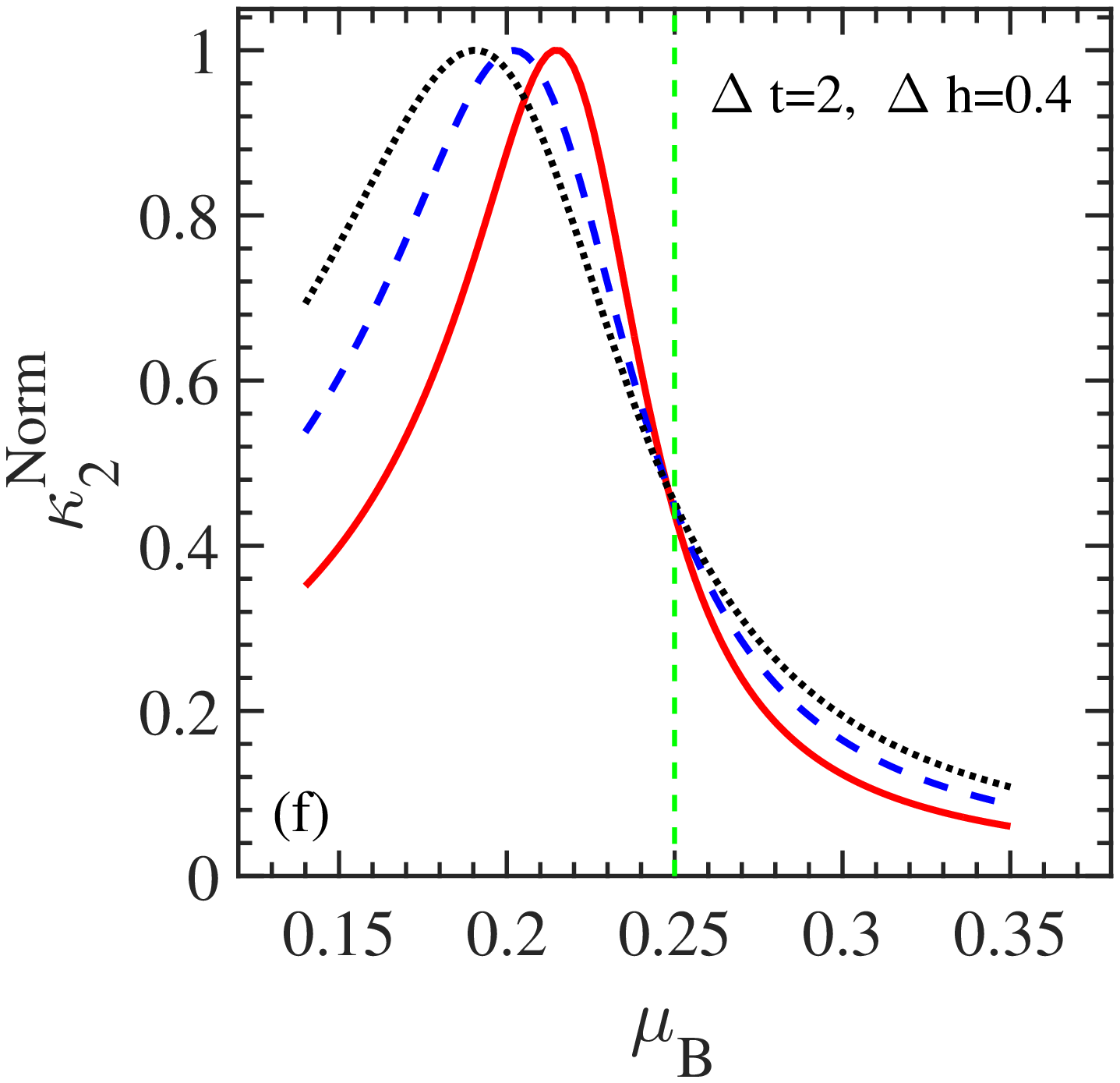}
    \includegraphics[width=0.31\textwidth]{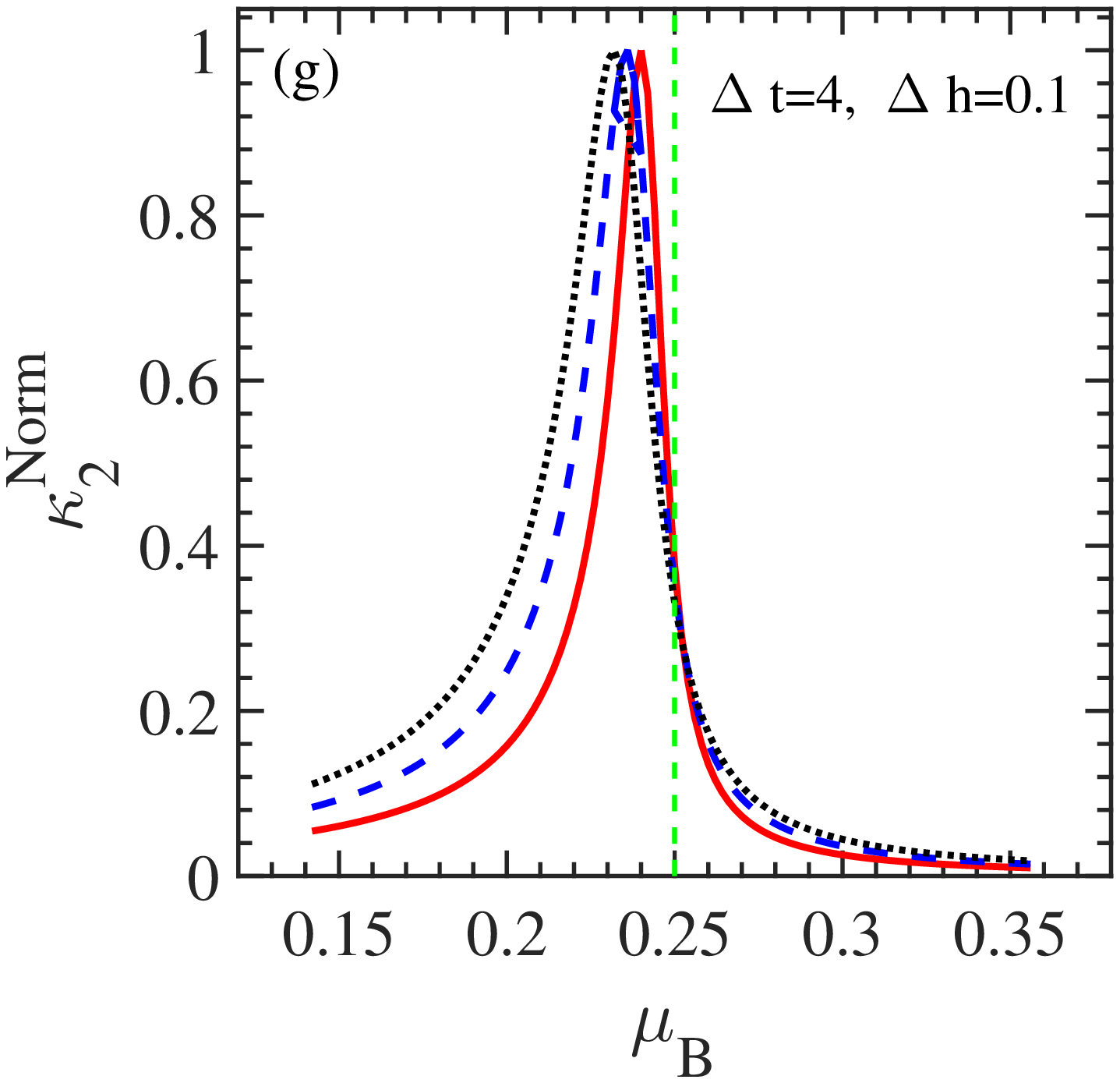}
    \includegraphics[width=0.31\textwidth]{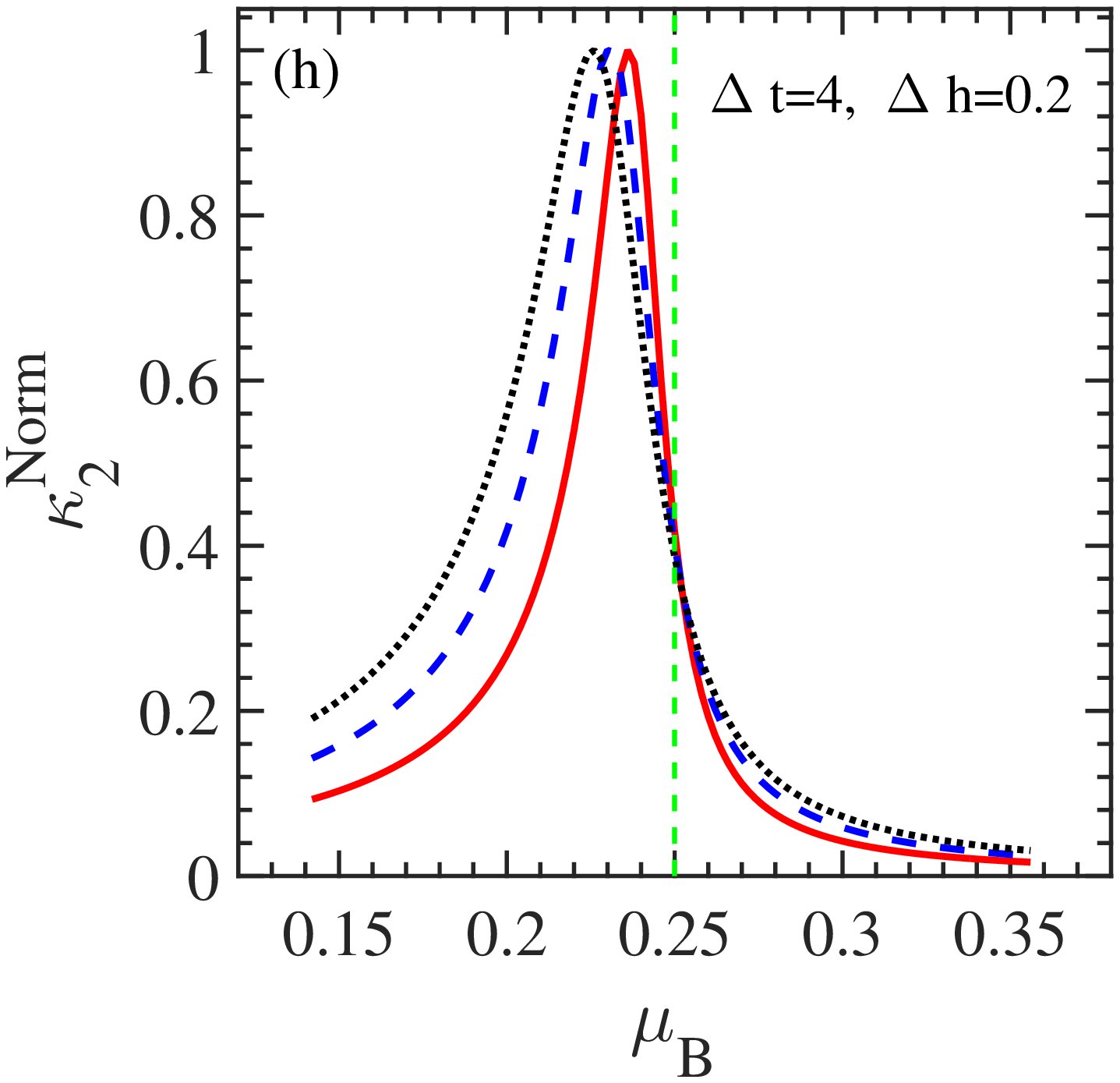}
    \includegraphics[width=0.31\textwidth]{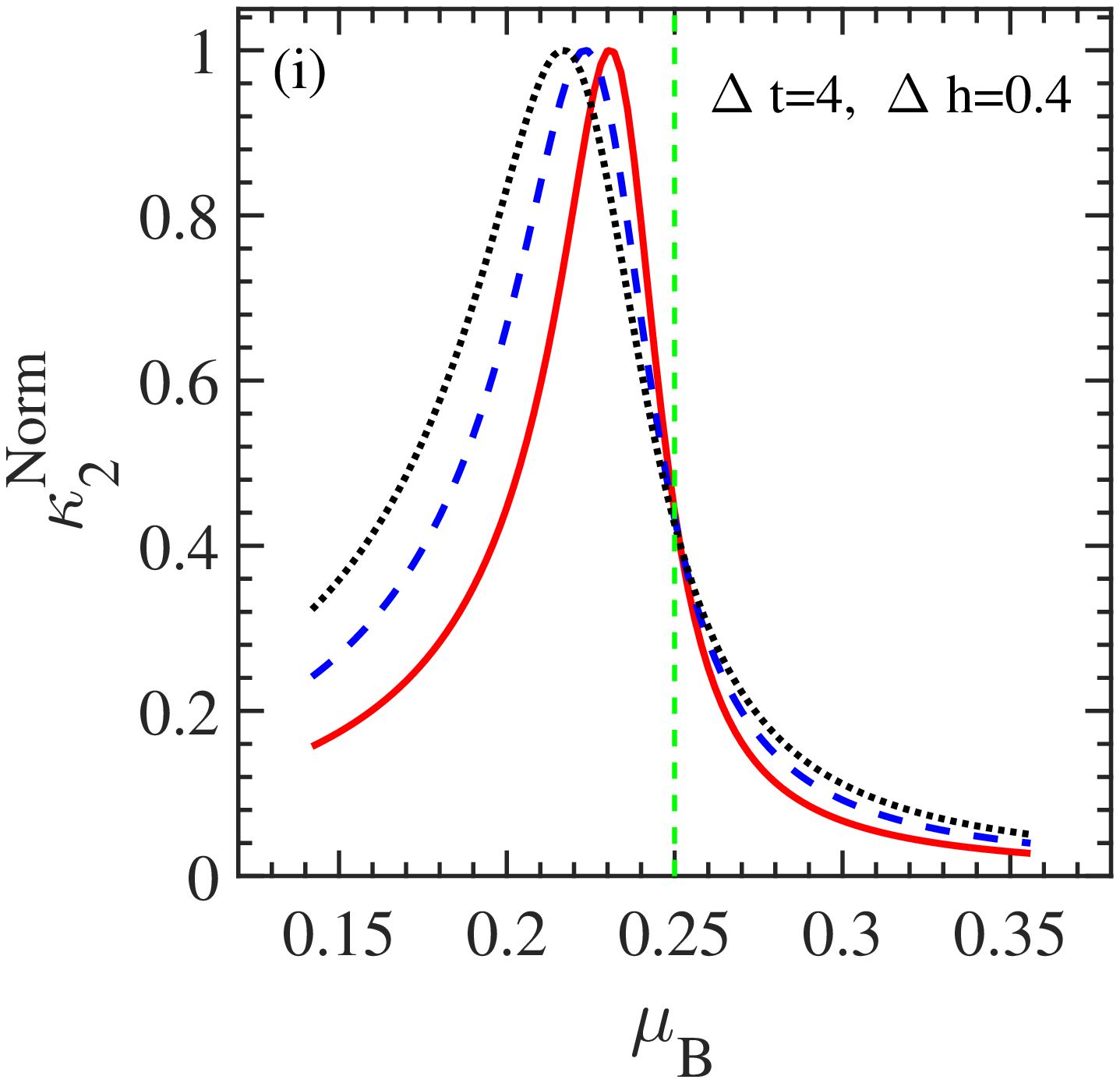}
    \caption{\label{Fig. 4}(Color online). Net-baryon chemical potential dependence of $\kappa_2^{Norm}$ at different $\Delta t$ and $\Delta h$ along three different freeze-out curves. $\Delta t = 1$, $2$ and $4$ for the up, middle and down panels, respectively. $\Delta h = 0.1$, $0.2$ and $0.4$ for the left, middle and right panels, respectively. The green dashed line shows the net-baryon chemical potential at the QCD critical point.}
\end{figure*}

To discuss the effects of $\Delta t$ and $\Delta h$ on the fixed point behavior in $\mu_B$ dependence of the normalized cumulants, $\Delta t$ is taken three different values $1$, $2$ and $4$. $\Delta h$ is also taken three different values $0.1$, $0.2$ and $0.4$. While the angles $\alpha_1$ and $\alpha_2$ are fixed at $4.5^\circ$ and $94.5^\circ$, respectively. The influence of the change of $\Delta t$ and $\Delta h$ on the fixed point behavior in $\kappa_2^{Norm}$, $\kappa_3^{Norm}$ and $\kappa_4^{Norm}$ are the same. For simplicity, let us take $\kappa_2^{Norm}$ as an example. The net-baryon chemical potential dependence of $\kappa_2^{Norm}$ is shown in Fig.~4. Results along the three freeze-out curves are still represented by the solid red curve, dashed blue curve and dotted black curve, respectively. The vertical dashed green line shows the critical net-baryon chemical potential.

In each column, the value of $\Delta h$ is fixed and the value of $\Delta t$ increases from top to bottom. It is clear that as $\Delta t$ gets bigger, the peak in $\kappa_2^{Norm}$ gets sharper. The position of fixed point shift to the bigger $\mu_B$ side a little. At the same time, the value of $\kappa_2^{Norm}$ at fixed point gets a little smaller. In each row, the value of $\Delta t$ is fixed while the value of $\Delta h$ increases from left to right. It is clear that the peak in $\kappa_2^{Norm}$ gets flatter as $\Delta h$ gets bigger. The position of fixed point shift to the smaller $\mu_B$ side a little. Meanwhile the value of $\kappa_2^{Norm}$ at fixed point gets a little bigger.

So the trend of the influence of $\Delta t$ and $\Delta h$ on the normalized cumulants and position of fixed point is just opposite. Smaller $\Delta t$ and bigger $\Delta h$ lead to flatter peak in $\mu_B$ dependence of $\kappa_2^{Norm}$ and the move of the fixed point to the smaller $\mu_B$ side, i.e. the peak of $\kappa_2^{Norm}$ in Fig.~4(c) is the flattest and the value of $\mu_B$ at the fixed point is the smallest. One more thing to note is that the position of the fixed point is all close to the vertical dashed green line in each sub-figure.

On the whole, the influence of $\Delta t$ and $\Delta h$ on the fixed point behavior is not so great as that of $\alpha_2$. In all case, the fixed point exists and is close to the critical net-baryon chemical potential when the $t$-direction and $h$-direction are nearly orthogonal.

Turning to the heavy-ion collision experiments, the energy ($\sqrt s$) dependence of the normalized cumulants can be got from the relation between $\mu_B$ and $\sqrt s$ given in Ref.~\cite{Phys.Rev.C.73.034905}, for details to see Ref.~\cite{cpc2022}.
It is easy to realize that the fixed point behavior also exists in the energy dependence of the normalized cumulants along the three different freeze-out curves as long as the $t$-direction and $h$-direction is not far from orthogonal to each other after mapped to the QCD $T-\mu_B$ plane. Based on the universality of the critical behavior, the fixed point behavior may be used to locate the QCD critical point in the relativistic heavy-ion collision experiments.

\section{Summary}

Through a linear mapping relation from the three-dimensional Ising model to QCD, the experimental freeze-out curves are mapped to the Ising $t-h$ plane. By using the parametric representation of the three-dimensional Ising model and the linear mapping relation, the net-baryon chemical potential dependence of normalized second- to fourth-order cumulants of the order parameter along three different freeze-out curves are studied.

Fixed point behavior is observed in the net-baryon chemical potential dependence of the normalized cumulants along three different freeze-out curves when the $h$-direction and $t$-direction is orthogonal or not far from orthogonal after mapping to the QCD $T-\mu_B$ phase plane. The fixed point is closer to the net-baryon chemical potential of QCD critical point than the peak of the cumulants.

The influence of three mapping parameters $\alpha_2$, $\Delta t$ and $\Delta h$, on the net-baryon chemical potential dependence of the normalized cumulants and the fixed point behavior is discussed. As the increase of $\alpha_2$, that is from $72^{\circ}$ to $117^{\circ}$ here, the net-baryon chemical potential at the fixed point increases, while the corresponding value of normalized cumulants decreases. When the $t$-direction and $h$-direction are nearly orthogonal to each other, the net-baryon chemical potential at the fixed point is much closer to the one at the QCD critical point.

The two parameters describing the width of the critical region of the three-dimensional Ising model, $\Delta t$ and $\Delta h$, have great influence on the behavior of cumulants, but little influence on the fixed point behavior. As the increase of $\Delta t$ and decrease of $\Delta h$, the peak in the net-baryon chemical potential dependence of the cumulants gets sharper, while the net-baryon chemical potential at the fixed point get a little bigger but not so obvious. And the values of $\mu_B$ at the fixed point are all close to that of $\mu_{Bc}$ when the $t$-direction and $h$-direction are orthogonal.

Fixed point behavior also exists in the energy dependence of the normalized cumulants along the experimental freeze-out curves when the $t$-direction and $h$-direction is orthogonal or not far from orthogonal after mapping to the QCD $T-\mu_B$ plane. It may be used to locate the QCD critical point in the relativistic heavy-ion collision experiments.

\ed